%% file: paper.tex
\documentclass[11pt]{article}
\pdfoutput=1
\usepackage{epsfig,sint,cite,amsfonts,amssymb,amsmath,bbold}
\usepackage[font=small,labelfont={sf,bf}]{caption}
\usepackage{subcaption}
\usepackage[T1]{fontenc}
\usepackage{tikz}
\usetikzlibrary{decorations.pathreplacing}
\usetikzlibrary{patterns} 
\usepackage[english]{babel}
\usepackage{simplewick}
\usepackage{dsfont}
\usepackage[utf8]{inputenc}
\usepackage{physics}
\usepackage{siunitx}
\usepackage{booktabs}
\input{macros.tex}

\begin{document}

\begin{titlepage}
\begin{flushright}
\hfill DESY-16-178\\
\end{flushright}

\vskip 0.5cm

\begin{center}

{\Large\bf A local factorization of the fermion determinant\\
in lattice QCD\\[0.5ex]} 

\end{center}
\vskip 0.25 cm
\begin{center}

{\large  Marco C\`e}
\vskip 0.125cm
Scuola Normale Superiore, Piazza dei Cavalieri 7, I-56126 Pisa, Italy\\
and INFN, Sezione di Pisa, Largo B. Pontecorvo 3, I-56127 Pisa, Italy\\
E-mail: marco.ce@sns.it\\
\vskip 0.5cm

{\large  Leonardo Giusti}
\vskip 0.125cm
Dipartimento di Fisica, Universit\`a di Milano--Bicocca,\\
and INFN, Sezione di Milano--Bicocca,\\
Piazza della Scienza 3, I-20126 Milano, Italy\\
E-mail: Leonardo.Giusti@mib.infn.it\\
\vskip 0.5cm

{\large  Stefan Schaefer}
\vskip 0.125cm
John von Neumann Institute for Computing (NIC),\\
DESY, Platanenallee 6, D-15738 Zeuthen, Germany\\
E-mail: Stefan.Schaefer@desy.de\\
\vskip 0.5cm

{\bf Abstract}
\vskip 0.35ex
\end{center}

\noindent
We introduce a factorization of the fermion determinant in lattice QCD with
Wilson-type fermions that leads to a bosonic action which is local in the block fields.
The interaction among gauge fields on distant blocks is
mediated by multiboson fields located on the boundaries of the blocks.
The resultant multiboson domain-decomposed hybrid Monte Carlo passes
extensive numerical tests carried out by measuring standard gluonic
observables. The combination of the determinant factorization
and of the one of the propagator, that we put forward recently, paves
the way for multilevel Monte Carlo integration in the presence of fermions. 
We test this possibility by computing the disconnected correlator of two
flavor-diagonal pseudoscalar densities, and we observe a significant
increase of the signal-to-noise ratio due to a two-level integration.
\vfill

\eject

\end{titlepage}

\section{Introduction}
State of the art algorithms for lattice QCD simulations require first integrating
out analytically the Grassmann quark fields, e.g.\ for two degenerate flavors, working with
the partition function
\be
Z = \int [dU]\, \{\det D \}^2 \,e^{-S_\mathrm{G}[U]}
\ee
with $D$ the massive Dirac operator and $S_\mathrm{G}$ the gauge action, and then
simulating this  effective gauge theory by Monte Carlo techniques. As a result,
the locality of the original action and of the observables is not manifest anymore,
because the fermion determinant and the propagator are nonlocal functionals of the
link variables $U$.

While necessary for making lattice QCD simulations feasible, the nonlocality
leads to severe limitations in practice. Local (link) update algorithms,
the method of choice for pure gauge theory, are not competitive anymore. The effective
gauge theory is instead simulated with variants of the {\it global} hybrid Monte Carlo (HMC)
algorithm~\cite{Duane:1987de}, with only its local variant performing comparably to other link update
techniques\footnote{Attempts to make proposals including dynamical fermions
based on link updates as in \Refs{Hasenbusch:1998yb,Knechtli:2003yt,Hasenfratz:2005tt} have not 
been adopted in large scale projects.}~\cite{Gehrmann:1999wr}.
For the same reason, noise reduction techniques based on the locality
of the theory, such as multihit or multilevel
algorithms~\cite{Parisi:1983hm,Luscher:2001up,Meyer:2002cd,DellaMorte:2007zz,
DellaMorte:2008jd,DellaMorte:2010yp}, have not yet been formulated
successfully in theories with fermions. They are expected to lead to an impressive
acceleration in those cases where the signal-to-noise ratio decreases exponentially
with the distance between the sources~\cite{Parisi:1983ae,Lepage:1989hd}.

Over the last two decades, there have been many attempts to rewrite the fermion determinant
via a local bosonic field theory. In the multiboson (MB) approach~\cite{Luscher:1993xx},
the bosonic action is ultralocal. The backreaction from the large number of
bosonic fields which are typically required, however, results in stiff gauge links and thus
in long autocorrelation times~\cite{Jegerlehner:1994cd}. In the domain-decomposed hybrid
Monte Carlo (DD-HMC), the determinants of the block Dirac operators
are factorized. The remainder, however, is not small, depends on the gauge
field values over the entire lattice, and needs to be represented by boson fields
with a nonlocal action~\cite{Luscher:2005rx}.

The aim of this paper is to introduce a factorization of the fermion determinant
in lattice QCD with Wilson-type fermions which can be represented by a bosonic
theory with a local action in the block gauge and pseudofermion fields. The first step
consists in factorizing out from the determinant the contribution depending
on gauge fields in distant blocks; see Eq.~(\ref{eq:step3}).
In the second step this factor, which deviates from the identity by terms suppressed as 
$\exp{-M_\pi \Delta}$ where $M_\pi$ is the pion mass
and $\Delta$ is the distance between the blocks, is taken exactly into account by introducing
multiboson fields on the boundaries of the blocks involved. As a result the final
bosonic action is local in the block fields.

Together with the factorization of the fermion observables presented
in Ref.~\cite{Ce:2016idq}, this opens the way for multilevel simulations of QCD.
We implement these ideas in a multiboson domain-decomposed hybrid Monte Carlo
(MB-DD-HMC), which we test extensively by measuring the two-point correlators of
the gluonic energy density, of the topological charge density, and of two flavor-diagonal
pseudoscalar densities. In all cases we observe a significant increase of the
signal-to-noise ratio due to a two-level integration.

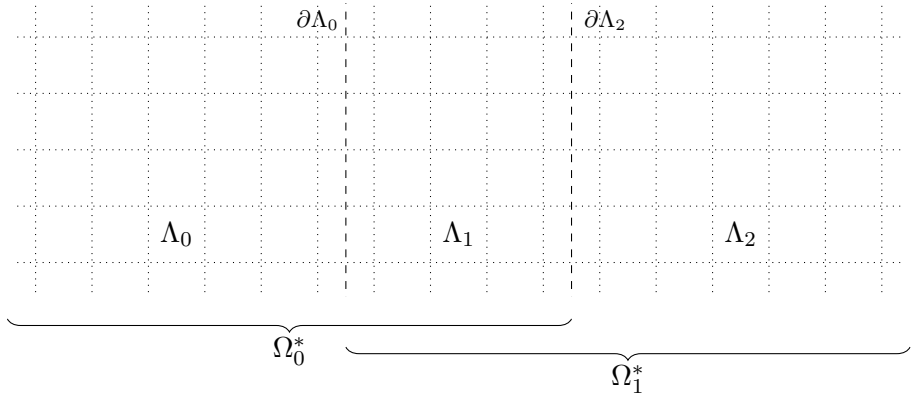
\begin{figure}[t!]
  \centering
  \begin{tikzpicture}[scale=1.5]
    \begin{scope}
      \clip (-0.2,-1.3) rectangle (7.7,+1.3);
      \draw[step=0.5, dotted] (-0.5,-2.5) grid (14,+2.5);
      \draw[dashed] (+2.75,-2.5) -- +(0,5);
      \draw[dashed] (+4.75,-2.5) -- +(0,5);
    \end{scope}
    \foreach \x in {0,1,2}
      \draw (2.5*\x+1.25, -0.75) node {$\Lambda_\x$};
    \foreach \x in {2}
      \draw (2.5*\x+0.05,1.15) node [font=\footnotesize] {$\partial\Lambda_\x$};
    \foreach \x in {0}
    \draw (2.5*\x+2.50,1.15) node [font=\footnotesize] {$\partial\Lambda_\x$};

\begin{scope}[xshift=-0.25cm, yshift=-1.50cm]
      \draw[decorate, decoration={brace, amplitude=5, mirror}] (0,0) -- +(5,0) node [midway, below=3] {$\Omega^*_{0}$};
    \end{scope}
    \begin{scope}[xshift=1.75cm, yshift=-1.75cm]
      \draw[decorate, decoration={brace, amplitude=5, mirror}] (1,0) -- +(5,0) node [midway, below=3] {$\Omega^*_{1}$};
    \end{scope}    
  \end{tikzpicture}
  \caption{Decomposition of the lattice in three thick time slices.}
  \label{fig:decomp}
\end{figure}

\section{Block decomposition of the determinant\label{sec:BD}}

The goal of the derivation in this and in the following section is a
decomposition of the effective fermion action in terms which are local in
the block gauge and scalar fields. The essential idea can be presented by
considering a decomposition of the lattice in three blocks
$\Lambda_i$, $i=0,1,2$, where each block may be the union of disconnected
regions, and the only requirement is that the Dirac operator does not connect
the blocks $\Lambda_0$ with $\Lambda_2$ directly. Without loss of generality,
we consider the simple case of a lattice with open boundary conditions in the
time direction and specify the blocks to be three thick
time slices; see Fig.~\ref{fig:decomp}. With minor modifications,
the same setup applies also to periodic boundary conditions. The general case,
in which more than three regions are considered, is given in
Appendix~\ref{a:evenodd}. With this domain decomposition,
the Hermitian $O(a)$-improved massive
Wilson-Dirac operator $Q=\gamma_5 D$ (see Appendix~\ref{app:Dw}) takes the block
form\footnote{The block terminology and decompositions used here follow
closely those in Ref.~\cite{Ce:2016idq}. To keep the notation compact,
a block matrix $Q_{\Lambda_{i,j}}$ denotes either a single block of
the matrix, or the full matrix with just that block different from zero.
Throughout the paper dimensionful
quantities are always expressed in units of the lattice spacing $a$, unless
explicitly specified.}
\begin{equation}
Q=\left ( 
\begin{matrix} 
Q_{\Lambda_{0,0}} & Q_{\Lambda_{0,1}}  & 0      \\
Q_{\Lambda_{1,0}} & Q_{\Lambda_{1,1}}  & Q_{\Lambda_{1,2}} \\
 0     & Q_{\Lambda_{2,1}}  & Q_{\Lambda_{2,2}} \\
\end{matrix}
\right )\; . 
\end{equation}
It is useful to define projection operators to the subspaces of quark fields supported
on the domains $\Lambda_i$ as
\be
   [P_{\Lambda_{i}}\psi](x) =
\begin{cases}
\psi(x) &  x\in \Lambda_i\, , \\[0.375cm]
0 & {\rm elsewhere}\; .
\end{cases}
\ee
In the following $P_{\Lambda_{i}}$ indicates the projector
irrespectively of the dimension of the full space on which it acts.
Following Ref.~\cite{Ce:2016idq}, we define the two-block operators
\begin{equation}
Q_{\Omega^*_i}=\left ( 
\begin{matrix} 
Q_{\Lambda_{i,i}} & Q_{\Lambda_{i,i+1}}  \\
Q_{\Lambda_{i+1,i}} & Q_{\Lambda_{i+1,i+1}} 
\end{matrix}
\right )\; ,
\end{equation}
where $\Omega^*_i=\Lambda_i\cup\Lambda_{i+1}$ and $i=0,1$.
The factorization of the determinant of $Q$ is achieved as described
in the following four steps.\\

\begin{description}
\item[Step 1] Introduce the two-block partitioning of the lattice
as defined in Appendix~\ref{app:DD} with $\Gamma=\Lambda_0\cup\Lambda_2$ and
$\Gamma^*=\Lambda_1$. Using Eq.~(\ref{eq:detblock}) the determinant 
can be factorized as
\begin{equation}
\begin{split}\label{eq:step1}
\hspace{-0.5cm}\det\, Q&=\det Q_{\Lambda_{1,1}} \det\, \left ( 
\begin{matrix} 
Q_{\Lambda_{0,0}}-Q_{\Lambda_{0,1}} Q_{\Lambda_{1,1}}^{-1} Q_{\Lambda_{1,0}} & -Q_{\Lambda_{0,1}} Q_{\Lambda_{1,1}}^{-1}
Q_{\Lambda_{1,2}} \\[0.25cm]
-Q_{\Lambda_{2,1}} Q_{\Lambda_{1,1}}^{-1} Q_{\Lambda_{1,0}}  & Q_{\Lambda_{2,2}} -Q_{\Lambda_{2,1}} Q_{\Lambda_{1,1}}^{-1} Q_{\Lambda_{1,2}} 
\end{matrix}
\right ) \,,
\end{split}
\end{equation}
where the second matrix on the rhs is the Schur complement associated to $\Gamma$.\\

\item[Step 2] Use again Eq.~(\ref{eq:detblock}) to obtain
\be\label{eq:step2}
\det\, Q = \frac{1}{\det\, Q^{-1}_{\Lambda_{1,1}} \det\left[P_{\Lambda_{2}}\,
Q^{-1}_{\Omega^*_1}\, P_{\Lambda_{2}}\right]
           \det\left[P_{\Lambda_{0}}\,  Q^{-1} \, P_{\Lambda_{0}}\right]}\,,
\ee
\end{description}
where each determinant is the one of the nonzero submatrix indicated; e.g.\
$\det[P_{\Lambda_{2}}\, Q^{-1}_{\Omega^*_1}\, P_{\Lambda_{2}}]$ stands for
the determinant of the inverse of the Schur complement of $Q^{-1}_{\Omega^*_1}$ in $\Lambda_2$
[see Eq.~(\ref{eq:Schur})]. For the first two determinants in the denominator, the
goal has been reached: they depend only on links from one or two time slices, respectively.

\begin{description}
\item[Step 3] Combine Eqs.~(\ref{eq:step1}), (\ref{eq:step2}) and
(\ref{eq:Scmpt2})
to rewrite the last determinant as
\be\label{eq:step3}
\hspace{-0.625cm}\frac{1}{\det\left[P_{\Lambda_{0}}\,  Q^{-1} \, P_{\Lambda_{0}}\right]} =
\frac{1}{\det\left[P_{\Lambda_{0}}\,  Q^{-1}_{\Omega^*_0}  \, P_{\Lambda_{0}}\right]}
\det\,
\left (
\begin{matrix} 
1 & P_{\Lambda_{0}} Q^{-1}_{\Omega^*_0} Q_{\Lambda_{1,2}} \\[0.25cm]
P_{\Lambda_{2}} Q^{-1}_{\Omega^*_1} Q_{\Lambda_{1,0}}  & 1
\end{matrix}
\right )\; . 
\ee
\end{description}
This step is suggested by the fact that for $x,y\in \Lambda_0$, the propagator
elements $Q^{-1}(x,y)$ are expected to be well approximated 
by the inverse of $Q_{\Omega^*_0}$ up to corrections suppressed
proportionally to $\exp{-M_\pi\Delta}$, where $M_\pi$ is the pion mass
and $\Delta$ is the thickness of the block $\Lambda_1$~\cite{Ce:2016idq},
see also below.\\

\begin{description}
\item[Step 4] Reduce the last determinant in Eq.~(\ref{eq:step3})
to the one of a matrix acting on one of the boundaries only. To this end,
Eq.~(\ref{eq:detblock}) is employed once more
\begin{equation}\label{eq:w}
\hspace{-0.8cm}\det\!\! 
\left (
\begin{matrix} 
1 & \!\!\!\!\!\!P_{\Lambda_{0}}\, Q^{-1}_{\Omega^*_0} Q_{\Lambda_{1,2}} \\[0.25cm]
P_{\Lambda_{2}}\, Q^{-1}_{\Omega^*_1} Q_{\Lambda_{1,0}}  & \!\!\!\!\!\!1
\end{matrix}
\right )
\!\!= \det (1 - P_{\partial\Lambda_{0}} Q^{-1}_{\Omega^*_0} Q_{\Lambda_{1,2}}   P_{\partial\Lambda_{2}} Q^{-1}_{\Omega^*_1}
Q_{\Lambda_{1,0}} P_{\partial\Lambda_{0}}),\!\!\!\\[0.375cm]
\end{equation}
where $P_{\partial\Lambda_{0}}$ and $P_{\partial\Lambda_{2}}$ are projectors on the inner boundary of
the thick time slices $\Lambda_0$ and $\Lambda_2$ respectively. They are
defined such that
\be
P_{\partial\Lambda_{i}} Q_{\Lambda_{i,j}} = Q_{\Lambda_{i,j}} P_{\partial\Lambda_{j}} = Q_{\Lambda_{i,j}} \; , \quad i\neq j\; .
\ee
In our thick time-slice partitioning the inner boundary of a block is the set of points at a distance $1$
from the previous and the next block. 
\end{description}
The factorized formula can finally be written as
\begin{equation}\label{eq:factfinal}
  \det\, Q= \frac{1}{\det\, Q^{-1}_{\Lambda_{1,1}}
    \det\left[P_{\Lambda_{0}}\,  Q^{-1}_{\Omega^*_0} \, P_{\Lambda_{0}}\right]
    \det\left[P_{\Lambda_{2}}\,  Q^{-1}_{\Omega^*_1} \, P_{\Lambda_{2}}\right]
    } \det\, (1-w)\; ,  
\end{equation}
where
\be
w=P_{\partial\Lambda_{0}} Q^{-1}_{\Omega^*_0}\, Q_{\Lambda_{1,2}}\, Q^{-1}_{\Omega^*_1}\,
Q_{\Lambda_{1,0}} \label{eq:wdef}
\ee
acts on the inner boundary field $P_{\partial\Lambda_{0}} \psi$. 

In Eq.~(\ref{eq:factfinal}) $\det Q^{-1}_{11}$ depends on the gauge field in the block $\Lambda_1$,
$\det\, [P_{\Lambda_{0}}\,  Q^{-1}_{\Omega^*_0} \, P_{\Lambda_{0}}]$ on the gauge fields in
$\Lambda_0\cup\Lambda_1$, and $\det\, [P_{\Lambda_{2}}\,  Q^{-1}_{\Omega^*_1} \, P_{\Lambda_{2}}]$
on the gauge field in $\Lambda_1\cup\Lambda_2$.  Only the (small) correction $\det\, (1-w)$
is a function of all the links of the lattice. Note that $\det (1-w)$ is real
since all other determinants entering Eq.~(\ref{eq:factfinal}) are real. 

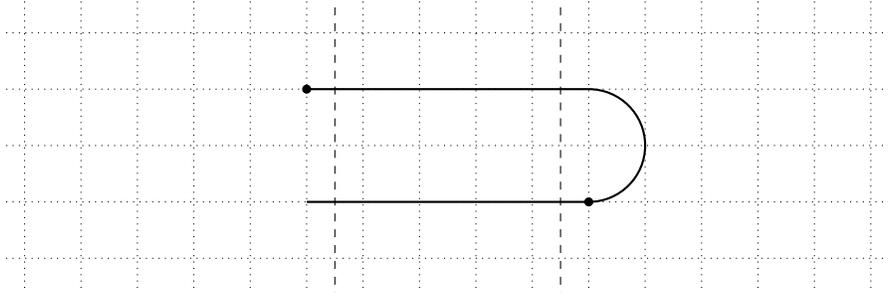
\begin{figure}[t!]
  \centering
  \begin{tikzpicture}[scale=1.5]
    \begin{scope}
      \clip (-0.2,-1.3) rectangle (7.7,+1.3);
      \draw[step=0.5, dotted] (-0.5,-2.5) grid (14,+2.5);
      \draw[dashed] (+2.75,-2.5) -- +(0,5);
      \draw[dashed] (+4.75,-2.5) -- +(0,5);
    \end{scope}
    \begin{scope}[thick, xshift=3.75cm]
      \draw[black] (-1.25,-0.5) -- +(+2.25,0);
      \draw[black] (+1.0,-0.5) -- +(0.25,0) arc [start angle=-90, end angle=90, radius=0.5] -- +(-2.5,0);
      \fill (-1.25,+0.5) circle (0.04);
      \fill (+1.25,-0.5) circle (0.04);
    \end{scope}
  \end{tikzpicture}
  \caption{Representation of the operator $w$. The black lines are full propagators while
    the thick dots are insertions of the effective hops; see Eqs.~(\ref{eq:effQ}).}
  \label{fig:operator_w}
\end{figure}

\subsection{Magnitude of $w$\label{s:magw}}
To shed some light on the size of the contributions of the global determinant $\det (1-w)$,
one can rewrite the matrix in terms of the full propagator $Q^{-1}$.
By considering two different partitions of the lattice in two blocks, first 
$\Lambda_{0}\cup \Lambda_1$ and $\Lambda_2$, and then  $\Lambda_0$ and $\Lambda_{1}\cup \Lambda_2$,
it is easy to show that 
\bea\label{eq:effQ}
P_{\partial\Lambda_{0}} Q^{-1}_{\Omega^*_0}\, Q_{\Lambda_{1,2}}  & = & P_{\partial\Lambda_{0}} Q^{-1}
\left\{Q_{\Lambda_{1,2}} + Q_{\Lambda_{2,1}}\,  Q^{-1}_{\Omega^*_0}\, Q_{\Lambda_{1,2}}\right\}\; , \nonumber\\[0.25cm]
P_{\partial\Lambda_{2}} Q^{-1}_{\Omega^*_1}\, Q_{\Lambda_{1,0}}  & = & P_{\partial\Lambda_{2}} Q^{-1}
\left\{Q_{\Lambda_{1,0}} + Q_{\Lambda_{0,1}}\, Q^{-1}_{\Omega^*_1}\, Q_{\Lambda_{1,0}}\right\}\; .
\eea
Therefore two propagators between the boundaries of the block
$\Lambda_0$ and $\Lambda_2$, each one multiplied by an effective
boundary operator, appear in the definition of $w$
(see Fig.~\ref{fig:operator_w}).
Numerical experience shows that, if the thickness $\Delta$ of the block
$\Lambda_1$ is large enough, each of these propagators will be suppressed
proportionally to $\exp{-M_\pi \Delta/2}$~\cite{Ce:2016idq}. Therefore
the norm of $w$ is expected to be suppressed as $\exp{-M_\pi \Delta}$.\\

\subsection{Spectrum of $w$\label{s:prop}}
Detailed knowledge of the spectrum of $(1-w)$ is required for the next step. 
According to Eqs.~(\ref{eq:wdef}) and (\ref{eq:Scmpt2}), the matrix $w$
can be written as a product of two Hermitian matrices 
\begin{equation}
w=[P_{\partial\Lambda_{0}}\, Q^{-1}_{\Omega^*_0}\, P_{\partial\Lambda_{0}}]
  [Q_{\Lambda_{0,1}} Q_{\Lambda_{1,1}}^{-1} Q_{\Lambda_{1,2}}\, Q^{-1}_{\Omega^*_1}\,
    Q_{\Lambda_{2,1}} Q_{\Lambda_{1,1}}^{-1} Q_{\Lambda_{1,0}}]
\end{equation}
acting on the interior boundary of the block $\Lambda_0$, which in turn
implies that $w$ is similar to $w^\dagger$~\cite{Radjavi:1969}. The characteristic
polynomial of $w$ has therefore real coefficients, and the complex eigenvalues $\delta_i$ come
in conjugate pairs. The spectrum of $w$ is symmetric with respect to the
real axis, and the determinant of $(1-w)$ is real as anticipated. As a
consequence of Sec.~\ref{s:magw}, the modulus of the eigenvalues $\delta_i$ 
are expected to be suppressed proportionally to  $\exp{-M_\pi \cdot \Delta}$.

\section{Multiboson factorization\label{sec:MBF}}
In the previous section, the goal of factorizing the determinant into
contributions which depend on the gauge field in the neighboring thick time slices
has almost been reached. Only the determinant of $(1-w)$ in \Eq{eq:factfinal} depends
on the gauge field over the whole lattice. For a suitably chosen thickness of the central
thick time slice, however, all eigenvalues of $w$ are expected to satisfy $|\delta_i|\ll 1$. This in turn
implies a large spectral gap for the matrix $(1-w)$, a fact which makes it possible to express its
determinant through a polynomial approximation of $(1-w)^{-1}$.

\subsection{Polynomial approximation}
As reviewed in Appendix~\ref{app:b}, a generalization of L\"uscher's original multiboson
proposal~\cite{Luscher:1993xx} to complex matrices~\cite{Borici:1995np,Borici:1995bk,Jegerlehner:1995wb} 
starts by approximating the function $1/z$, with $z\in\mathbb{C}$, by the polynomial
\be\label{eq:bella}
P_N(z) \equiv \frac{1-R_{N+1}(z)}{z} = c_N \prod_{k=1}^N (z-z_k) \;,
\ee
where $N$ is chosen to be even, and the $N$ roots of $P_N(z)$ are obtained by requiring that
for the remainder polynomial $R_{N+1}$ holds $R_{N+1}(0)=1$. The roots $z_k$ can be chosen to lie
on an ellipse passing through the origin of the complex plane with center $1$ and foci
$1 \pm c$ (see Appendix~\ref{app:b}),
\begin{equation}
u_k = 1- z_k =\cos{\left(\frac{2\pi k}{N+1}\right)} + i \sqrt{1-c^2} \sin{\left(\frac{2\pi k}{N+1}\right)}\; ,
\quad k=1,\dots,N  \;.
\end{equation}

\subsection{Approximation of the determinant}
This polynomial can be used to approximate the inverse determinant 
\be\label{eq:bella2}
\det (1-w)\, \det\{ P_N(1-w)\} = \det\{1-R_{N+1}(1-w)\}\; ,
\ee
where, if for all eigenvalues of $w$ holds  $|\delta_i|<1$,  the rhs side converges exponentially to 1 as $N$ is increased.
Since $w$ is similar to $w^\dagger$, and the $u_k$ come in complex conjugate pairs,
the approximate determinant can be written in a manifestly positive form,
\begin{equation}
\det\{ P_N(1-w)\}^{-1} =  C \prod_{k=1}^{N/2} {\det}^{-1} \big\{ (u_k -w )^\dagger (u_k -w) \big\} \\[0.25cm]
 = C \prod_{k=1}^{N/2} {\det}^{-1} ( W_{\sqrt{u_k}}^\dagger \, W_{\sqrt{u_k}})
\label{e:pV}
\end{equation}
with an irrelevant constant $C$ and 
\be\label{eq:Wz}
W_z = \begin{pmatrix} 
 z \, P_{\partial\Lambda_{0}} & P_{\partial\Lambda_{0}} Q^{-1}_{\Omega^*_0} Q_{\Lambda_{1,2}}  \\[0.25cm]
  P_{\partial\Lambda_{2}} Q^{-1}_{\Omega^*_1} Q_{\Lambda_{1,0}}  & z \, P_{\partial\Lambda_{2}} 
  \end{pmatrix}\; . 
\ee
In the last equality of \Eq{e:pV}, the reverse substitution of the one in \Eq{eq:w}
has been performed.
For the determination of the approximation, it is advantageous to work with the
operator $w$ (acting on $\partial\Lambda_{0}$ only) since the order of the polynomial
is reduced  by about a factor of 
$2$ for a given accuracy. The expression (\ref{e:pV}) with $W_z$ acting on
$\partial\Lambda_{0}$ and  $\partial\Lambda_{2}$, however, 
allows in the next step for a fully factorized domain decomposition
of the fermion action.

\subsection{Multiboson action}
For two flavors of quarks we can finally represent the determinants by scalar fields~\cite{Weingarten:1980hx}\footnote{The identity
$\det Q^{-1}_{\Lambda_{1,1}} \cdot \det\,  [P_{\Lambda_{0}} Q^{-1}_{\Omega^*_0} P_{\Lambda_{0}}] =
\det\, Q^{-1}_{\Omega^*_0}$ can be used to speed up the simulation when region $1$ is active.}
\begin{equation}\label{eq:act}
\begin{split}
& \frac{\det Q^2}{\det\{1-R_{N+1}(1-w)\}^2} =\frac{1}{
    \det\, [Q_{\Lambda_{1,1}}^{-1}]^2 \cdot \det\,  [P_{\Lambda_{0}} Q^{-1}_{\Omega^*_0} P_{\Lambda_{0}}]^{2} \cdot
\det\, [P_{\Lambda_{2}}  Q^{-1}_{\Omega^*_1}P_{\Lambda_{2}}]^{2} } \times\\[0.25cm]
& \times \det\,\{P_N(1-w)\}^{-2} = \, C' \int [d\phi_0d\phi_0^\dagger] e^{-|P_{\Lambda_{0}}  Q^{-1}_{\Omega^*_0} \phi_0|^2 }
 \int [d\phi_1d\phi_1^\dagger] e^{-|Q_{\Lambda_{1,1}}^{-1}\phi_1|^2 }\cdot\\
      & \int [d\phi_2 d\phi_2^\dagger] e^{-|P_{\Lambda_{2}}  Q^{-1}_{\Omega^*_1} \phi_2|^2 }\cdot 
       \prod_{k=1}^{N} \left \{
       \int [d\chi_k d\chi_k^\dagger] e^{-|W_{\sqrt{u_k}} \chi_k|^2 } \right \}\; , \\
\end{split}
\end{equation}
where $C'$ is another irrelevant numerical constant.
Each scalar field $\phi_i$ is confined to the corresponding region $\Lambda_i$, $i=0,1,2$.
The $N$ fields $\chi_k$ live on the outer boundaries of region $\Lambda_1$. We can
decompose them as
$\chi_k=\eta_k+\xi_k$, with  $\eta_k=P_{\partial\Lambda_{0}} \chi_k$ and $\xi_k=P_{\partial\Lambda_{2}} \chi_k$,
and split explicitly the contributions from the inner boundaries of regions $\Lambda_0$ and
$\Lambda_2$ as 
\begin{equation}
\label{eq:multiboson}
\begin{split}
|W_{z} \chi_k|^2 &=
|z|^2 |\eta_k|^2 + |z|^2 |\xi_k|^2+|P_{\partial\Lambda_{2}} Q^{-1}_{\Omega^*_1} Q_{\Lambda_{1,0}} \eta_k|^2+|P_{\partial\Lambda_{0}} Q^{-1}_{\Omega^*_0}
Q_{\Lambda_{1,2}} \xi_k|^2\\
&+\big[z (\xi_k,Q_{\Lambda_{2,1}} Q^{-1}_{\Omega^*_0} \eta_k) +z^*\, (\xi_k, Q^{-1}_{\Omega^*_1} Q_{\Lambda_{1,0}} \eta_k)
  + \text{c.c.}\big]\; ,
\end{split}
\end{equation}
{\it The dependence of the bosonic action from the gauge field in block $\Lambda_0$ and
$\Lambda_2$ is thus factorized}. Interestingly, the terms in Eq.~(\ref{eq:multiboson})
which will contribute to the forces in region $\Lambda_0$ always start (or end) on the inner
boundary of $\Lambda_2$ and vice versa. The matrices in Eq.~(\ref{eq:multiboson})
contain one boundary to boundary quark propagator which is suppressed
exponentially in $\Delta$, see Eq.~(\ref{eq:effQ}), and so do the corresponding forces.

\subsection{Order of the polynomial}
The order of the polynomial can be fixed, for the required precision,
by employing Eq.~(\ref{eq:Rnp1}) in Appendix~\ref{app:b}. This
guarantees that
\be\displaystyle
{\rm max}_{||v||=1}\, ||\,[1-(1-w)P_N(1-w)\,]\, v||\leq {\rm max}_i\, |\delta_i|^{N+1} = |\delta|^{N+1}_{\rm max}
\ee
where $v$ is a generic vector on which $w$ acts. Then Eq.~(\ref{eq:bella2}),
when $|\delta|^{N+1}_{\rm max}\ll 1$ and $\Tr R_{N+1}(1-w)\ll 1$, implies
\be
\det (1-w)\, \det\{ P_N(1-w)\} = 1 - \Tr R_{N+1}(1-w) +\dots \; .  
\ee
At the first order in the expansion, the relative error which
one makes on the determinant is therefore
\be\label{eq:boundD}
|\Tr R_{N+1}(1-w)| \leq  \sum_i |\delta_i|^{N+1} \leq
\sum_{i=1}^{N_{\rm ev}} |\delta_i|^{N+1} + (6 L^3-N_{\rm ev})|\delta_{N_{\rm ev}+1}|^{N+1} \,,
\ee
where in the last inequality the contribution from the $N_{\rm ev}$ eigenvalues with the
highest modules, i.e.\ $|\delta_i|$ sorted decreasingly,
has been treated separately and $L$ is the spatial length in lattice units.
If most of the modes have a modulus significantly smaller than $|\delta|_{\rm max}$ and if $N$ is
large, the sum on the rhs of Eq.~(\ref{eq:boundD}) will not generate a large factor; see below.
Given the distribution of the eigenvalues of $w$, the circle centered in
$1$ with radius $1$ is a natural choice for the polynomial
approximating $(1-w)^{-1}$ that we adopt in the following. However one could
optimize further the approximation by working with an ellipse, and tuning the value
of $c$.

\subsection{Reweighting factor}
A given correlation function of a string of fields $O$ can finally be written as 
\bea\label{eq:rwgt}
\langle O \rangle  = \frac{\langle O\, {\cal W}_N \rangle_N}{\langle {\cal W}_N \rangle_N}
& = &
\frac{\langle O_{\rm fact}\, \rangle_N}{\langle {\cal W}_N \rangle_N} +
\frac{\langle O\,  {\cal W}_N - O_{\rm fact} \rangle_N}{\langle {\cal W}_N \rangle_N}\; ,
\eea
where $O_{\rm fact}$ is a rather precise factorized approximation of
$O$ (see Ref.~\cite{Ce:2016idq} for
instance) and $\langle\cdot\rangle_N$ indicates the expectation value in the theory defined by
the multiboson action at finite $N$. Since both the action and the observable are
factorized, the expectation value $\langle O_{\rm fact}\, \rangle_N$ can be computed with
a multilevel algorithm by generating gauge field configurations with the multiboson action at
finite $N$. All other quantities in Eq.~($\ref{eq:rwgt}$) can be computed with a one-level
Monte Carlo procedure. For two flavors, the reweighting factor ${\cal W}_N$ is
\be
{\cal W}_N = \det\{1-R_{N+1}(1-w)\}^2\,. 
\ee
This expression is easily evaluated as 
\be\label{eq:WNeta}
{\cal W}_N = \frac{\int [d \eta] [d\eta^\dagger]
e^{-|(1-R_{N+1})^{-1} \eta|^2}}{\int [d \eta] [d\eta^\dagger] e^{-\eta^\dagger \eta}}\; ,
\ee
where the exponent can be computed by a Taylor expansion,
and as usual the integral over $\eta$ can be replaced by random samples. For the special case of
a circle, the simplification $R_{N+1}(1-w)=w^{N+1}$ applies.

\begin{figure}[t!]
  \centering
  \includegraphics[width=0.49\columnwidth]{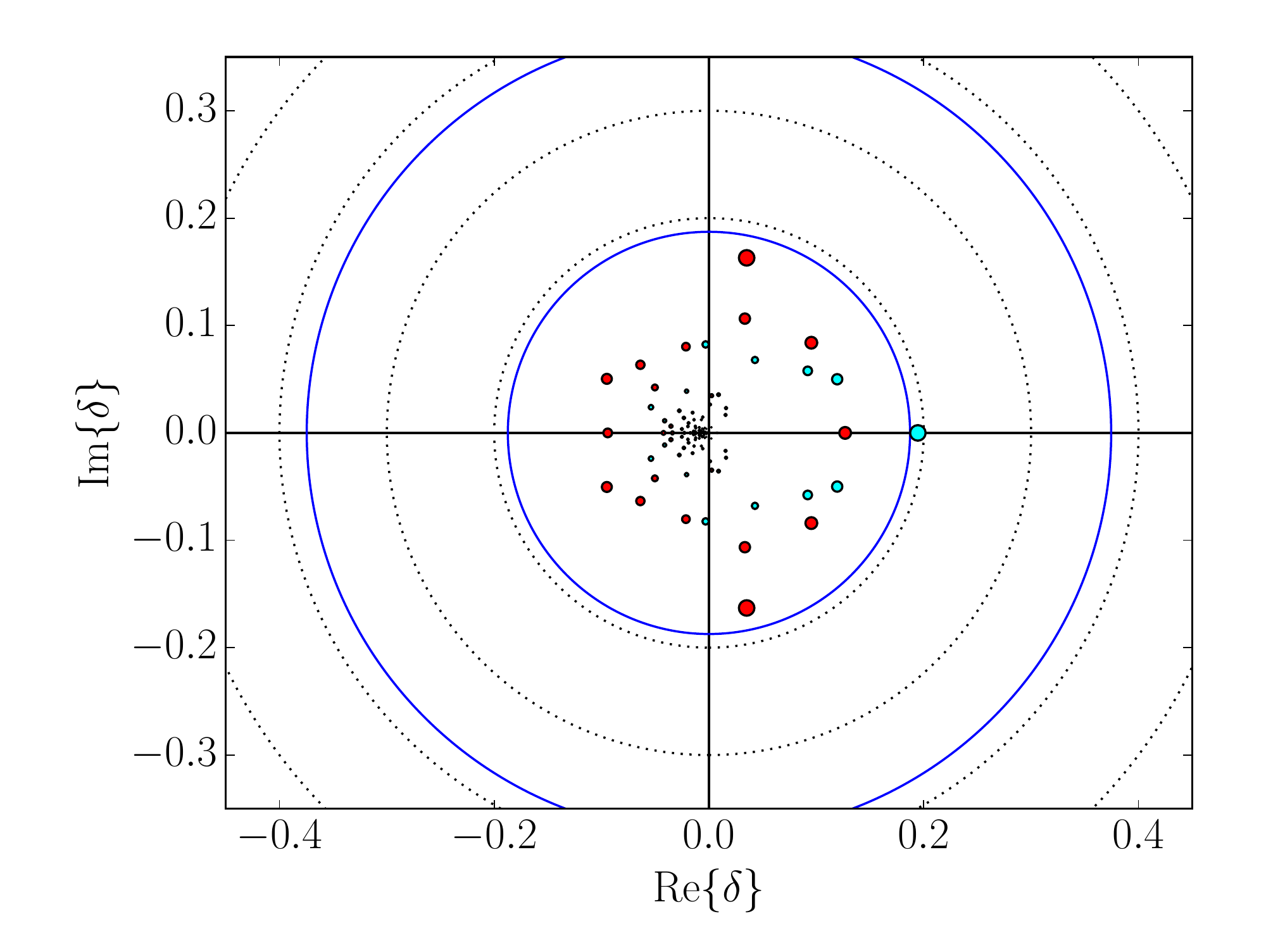}
  \includegraphics[width=0.49\columnwidth]{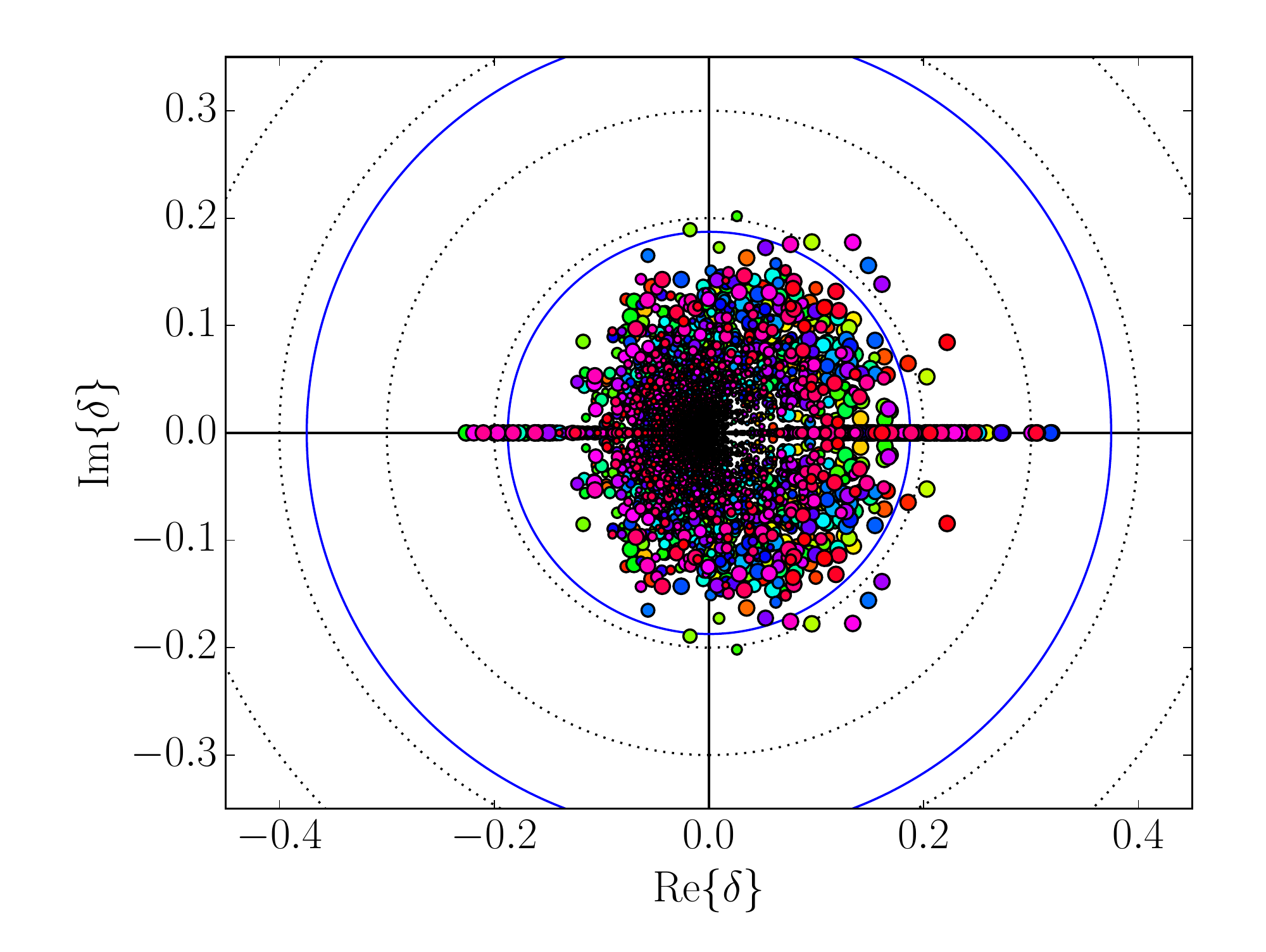}
  \caption{Left: Eigenvalues $\delta_i$ of $w$ ($\Delta=12$) with the largest norm for
    two typical configurations. Right: As in the left panel but for all $200$ configurations.
    In both panels the blue circles have radius $\bar\delta=\exp{-M_\pi \Delta}$ and $2\, \bar\delta$.}
\label{fig:spectrum_2cfg}
\end{figure}

\section{Numerical tests on the spectrum of $w$\label{s:numw}}
The feasibility of the whole proposal hinges
crucially on the assumption that the spectrum of the operator $(1-w)$ is confined into a disk
around $1$
in the complex plane, with a radius significantly below unity. Only in this case, a small number
of bosonic fields $N$ in \Eq{eq:multiboson} leads to a good enough approximation at reasonable
computational cost.

To test this assumption, we have generated a set of $200$ configurations 
with the Wilson gluonic action and with two flavors of nonperturbatively $O(a)$-improved Wilson quarks as 
defined in Appendix~\ref{app:Dw}, with
$\beta=5.3$, $c_{_{\rm SW}}=1.90952$, $c_{_{\rm F}}=c'_{_{\rm F}}=1$, $\kappa=0.13625$, $T\times L^3=64\times 32^3$
and open boundary conditions. The lattice spacing is $0.0652(6)\,\fm$, while the pion mass in
lattice units is $M_\pi=0.1454(5)$ corresponding to $440(5)$~MeV \cite{Fritzsch:2012wq}.

For $\Delta=8,12$ and $16$, we have computed with the Arnoldi algorithm the $60$ approximate eigenvalues
$\delta_i$ of $w$ with the largest absolute value. In the left plot of Fig.~\ref{fig:spectrum_2cfg},
they are shown for
$\Delta=12$ and on two typical configurations. As expected, the eigenvalues are either real or
appear in complex conjugate pairs. For one configuration (green points) the eigenvalue with the largest
absolute value is real, while for the other one (red points) two eigenvalues with opposite imaginary
parts have the largest absolute value. We find that both possibilities are common;
see the last column of Table~\ref{tab:spectrum}. In the right plot of Fig.~\ref{fig:spectrum_2cfg} we
show the eigenvalues $\delta_i$ again for $\Delta=12$ but for all configurations. The blue circles in these plots
have radius $\bar \delta$ and $2\, \bar\delta$, where
\begin{equation}
\bar \delta = \exp{-M_\pi \Delta} \;.
\end{equation}
\begin{figure}[t!]
  \centering
  \includegraphics[width=0.49\columnwidth]{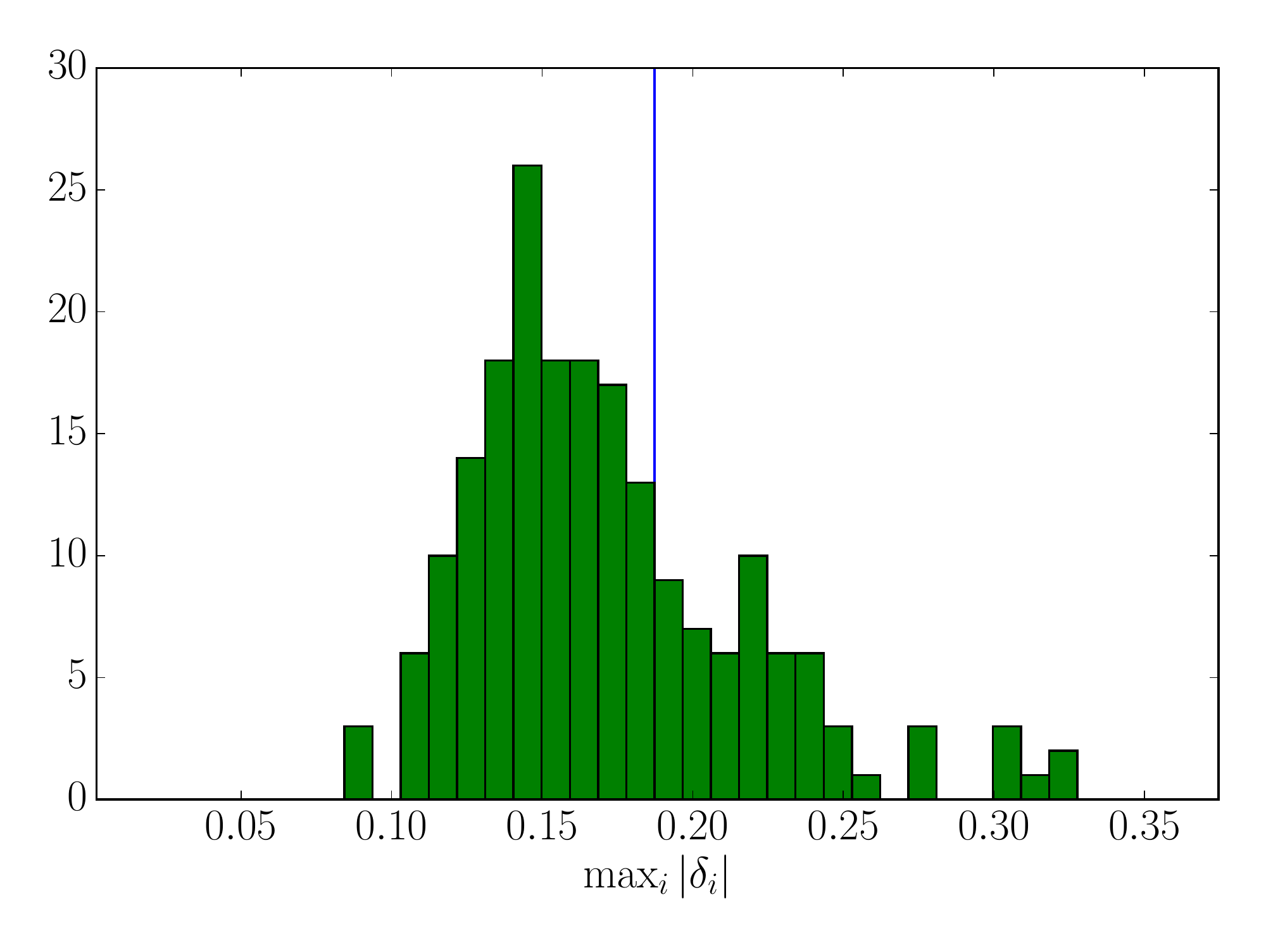}
  \includegraphics[width=0.49\columnwidth]{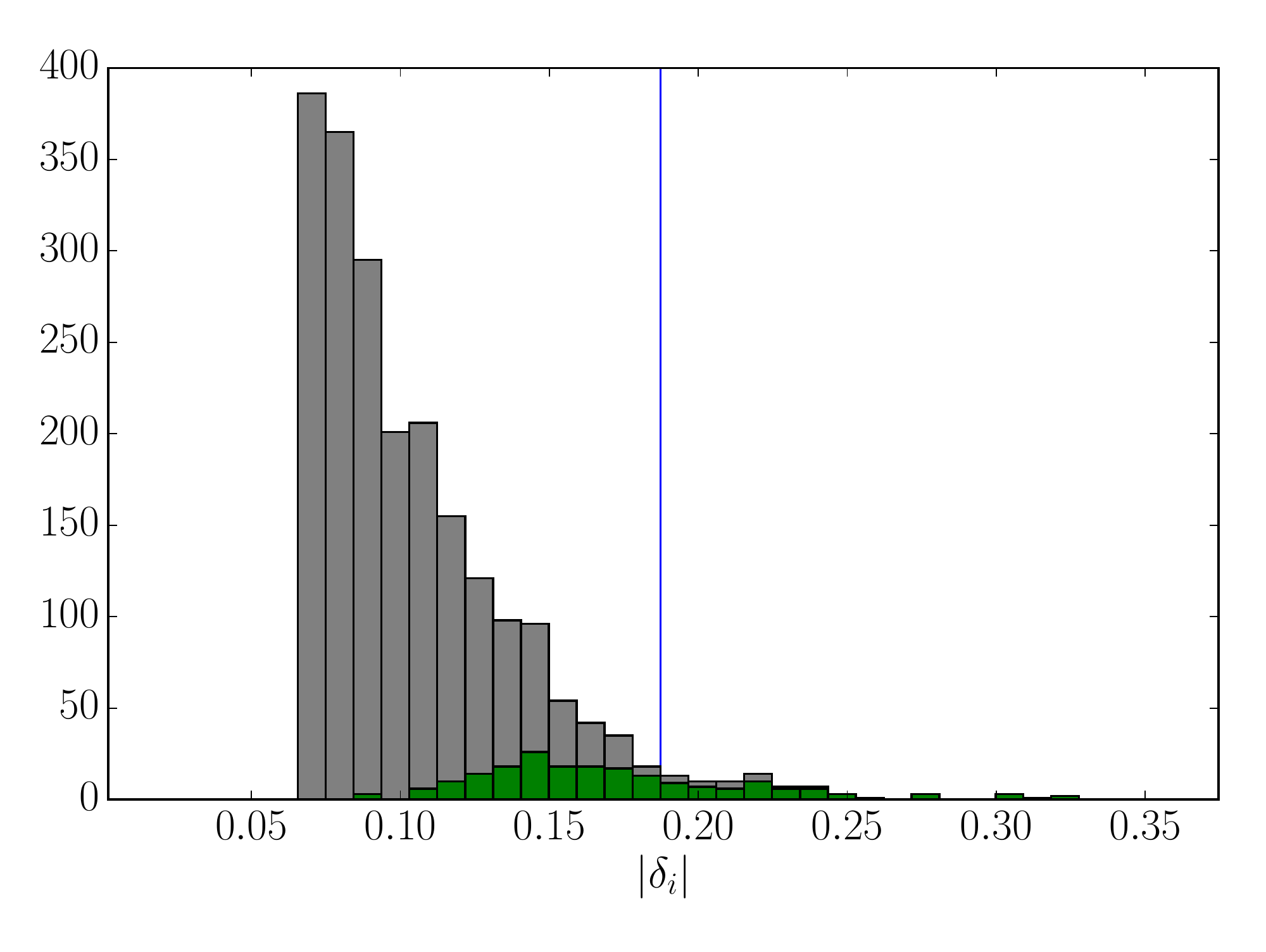}
  \caption{Left: Distribution of the eigenvalue of $w$ ($\Delta=12$) with the
largest absolute value. The vertical blue line is at $\abs{\delta_i}=\bar\delta$.
Right: As in the left panel but for the eigenvalues with $|\delta_i|>0.35\,\bar{\delta}$.}
  \label{fig:hist_delta}
\end{figure}
\begin{table}[t]
  \centering
  \caption{Properties of the spectrum of $w$ for different values of
$\Delta$. $f_{\Re}$ is the fraction of configurations for which $\delta_i$ with the largest absolute
value is real.}
\label{tab:spectrum}
\begin{tabular}{S[table-format=2]S[table-format=3]S[table-format=1.4]S[table-format=1.4]S[table-format=1.4]S[table-format=1.4]S[table-format=2.1]}
    \toprule
{$\Delta$} & {\texttt{ncfg}} & {$\bar\delta$} & {$\ev{\max_i\abs{\delta_i}}$} &
{$\sigma(\max_i\abs{\delta_i})$} &
 {$\max\max_i\abs{\delta_i}$} & {$f_{\Re} (\si{\percent})$} \\
    \midrule
        8 & 200 & 0.3273 & 0.2886 & 0.0616 & 0.5130 & 48.5 \\
       12 & 200 & 0.1710 & 0.1692 & 0.0453 & 0.3193 & 46.5 \\
       16 & 200 & 0.1072 & 0.0951 & 0.0284 & 0.1977 & 45.5 \\
    \bottomrule
  \end{tabular}
\end{table}
The distribution of the eigenvalue with the largest magnitude is
shown in the left plot of Fig.~\ref{fig:hist_delta}. It is peaked at a value slightly
smaller than $\bar\delta$, denoted by a vertical blue line, and extends up to $\sim 2\bar\delta$.
The results for the largest eigenvalue norm computed over the $200$ configurations, its
average value and the estimate of its standard deviation
are also reported in Table~\ref{tab:spectrum}. In the right plot of Fig.~\ref{fig:hist_delta}
we also report the distribution of the absolute value of the eigenvalues limited to those
with $\abs{\delta_i}>0.35\,\bar{\delta}$. 

A clear message emerges from these data. The largest eigenvalue of the relevant operator $w$ decreases
proportionally to $\exp{-M_\pi \Delta}$, with a prefactor of order $1$. This in turn implies that
$(1-w)$ has a large gap if $\Delta$ is properly tuned. The relative error on the determinant
computed as in Eq.~(\ref{eq:boundD}) at various values of $N$ compares well with
$|\delta|^{N+1}_{\rm max}$ configuration by configuration.  No big prefactors appear because the
eigenvalues do not accumulate near the maximum one, and the approximation gets exponentially more
precise toward the center of the circle. We have also computed the
reweighting factor as defined in Eq.~(\ref{eq:WNeta}). Its value, again for $N=12$ and estimated
with 4 random sources per configuration, deviates from $1$ by at most $4.5\cdot 10^{-6}$ again
in line with the expectation. At the level of precision of most contemporary
simulations the impact of the reweighting factor is therefore negligible.

\section{Numerical implementation of MB-DD-HMC}
The effective action in \Eq{eq:act} can be simulated using variants of the hybrid Monte Carlo
algorithm~\cite{Duane:1987de}. The implementation does not pose particular problems. To define
the setup for the tests discussed below, we mention a few essential points only. In the following
we distinguish between two basic contributions: the determinants of the block operators
such as $\det\,  [P_{\Lambda_{0}} Q^{-1}_{\Omega^*_0} P_{\Lambda_{0}}]$, and the multiboson contributions
responsible for the coupling between the blocks $\Lambda_{0}$ and $\Lambda_{2}$.

\subsection{Block action}
In the denominator of  \Eq{eq:act}, there are three determinants derived from the 
block decomposition of the fermion determinant. For the sake of the presentation we focus on
$\det\,  [P_{\Lambda_{0}} Q^{-1}_{\Omega^*_0} P_{\Lambda_{0}}]$, with the other two being treated analogously.
Since in general the thick time slices are not particularly thin, further decomposition of this determinant 
is necessary in order to have a cost-efficient simulation. One possibility is to apply 
mass preconditioning~\cite{Hasenbusch:2001ne,Hasenbusch:2002ai}, that is to write
\be
\label{eq:hb}
\det\,  [P_{\Lambda_0} Q^{-1}_{\Omega^*_0} P_{\Lambda_0}] 
 =\prod_{i=1}^{N_\mu-1} \frac{\det\,  [ P_{\Lambda_0} \{Q_{\Omega^*_0}(\mu_i)\}^{-1} P_{\Lambda_0} ]  }
{\det\,  [P_{\Lambda_0}  \{Q_{\Omega^*_0}(\mu_{i+1})\}^{-1} P_{\Lambda_0}]  }
\cdot
 \det\,   [P_{\Lambda_0}  \{ Q_{\Omega^*_0}(\mu_{N_\mu})\}^{-1} P_{\Lambda_0}]  
\ee
with $0=\mu_1<\mu_2<\dots<\mu_{N_\mu}$ and  $Q_{\Omega^*_0}(\mu)=Q_{\Omega^*_0}+i\mu P_{\Lambda_0}$.
The pseudofermion heatbath is then performed using 
\be
[P_{\Lambda_0}\{Q_{\Omega^*_0}(\mu)\}^{-1} P_{\Lambda_0}]^{-1}
=Q_{\Lambda_{0,0}}-Q_{\Lambda_{0,1}}Q_{\Lambda_{1,1}}^{-1}Q_{\Lambda_{1,0}}+i\mu P_{\Lambda_0}\,.
\ee
In the numerical tests described in the following we have used $N_\mu=5$, with twisted-mass values
$\mu_i=0.0, 0.001, 0.005, 0.1, 0.5$ for $i=1,\dots,5$.

\subsection{Multiboson action}
The contribution of the multiboson fields is by construction small and therefore
preconditioning does not seem to be necessary in a first implementation. The computation
of the forces themselves is straightforward, while the heatbath for the bosonic fields
requires further attention.

For the fields $\chi_k$, distributed according to the action
$S_{\mathrm{b},k}=|W_{\sqrt{u_k}} \chi_k|^2 $, the heatbath can be performed in
the usual fashion, by acting with the inverse of $W_{\sqrt{u_k}}$ on Gaussian
random fields located on the inner boundaries of regions $\Lambda_0$ and $\Lambda_2$. One
way to solve the corresponding linear system is discussed in Appendix \ref{a:invW}.
The cost of these inversions is negligible compared to the one of the molecular dynamics
evolution.
In the numerical tests presented in the following we have used $N=12$ multiboson fields
$\chi_k$, with the roots $z_k$ chosen to lie on the circle of  radius $1$ centered in $1$.

\section{Numerical tests of MB-DD-HMC}
In order to test the potentiality of the two-level MB-DD-HMC algorithm
described in the previous section, we have taken a subset of $n_0=32$
configurations spaced by at least $80$ molecular dynamics units (MDUs)  among the
$200$ described in Sec.~\Sect{s:numw}, which we can safely assume to be
independent.\footnote{Since with $N=12$ the reweighting factor is negligible within
the statistical precision of our observables, for testing purposes it is appropriate to use the
level-0 configurations already generated with the exact action.} Starting from
each of them, we have generated $n_1=45$ level-1 configurations spaced by $4$
MDUs by keeping fixed the spatial links on the boundaries $\partial\Lambda_{0}$
and $\partial\Lambda_{2}$  and all the links in between. The region $\Lambda_1$ extends
between time slices $24$ and $35$, corresponding to a thickness of
$\Delta\approx0.8$\,fm and $M_\pi\Delta \approx 1.7$. 

For the gauge variables in each of the two active regions $\Lambda_0$ and
$\Lambda_2$, the molecular dynamics in the HMC can be integrated with the
following nested three-level scheme. The forces derived from the multiboson
fields are integrated on the outermost level with a second order
Omelyan-Mryglod-Folk (OMF)~\cite{Omelyan2003272} integrator, with 12 steps
per trajectory of length 2.
On the second level, all forces derived from the block determinants in
\Eq{eq:hb} are integrated, with one step of the fourth order OMF integrator per
outer step. The third level, which consists again of one  fourth order OMF step,
takes care of the gauge forces. This scheme, which is very similar to the ones
used in \Ref{Luscher:2012av}, leads to an acceptance rate of 94\%.

\subsection{Correlation functions of gluonic operators}
The primary local gluonic observables that we measure to test the algorithm are 
the energy  and the topological charge densities summed
over the time slices, i.e. 
\bea
\bar e (x_0) = \frac{1}{4} \sum_{\vec x} F_{\mu\nu}^{a}(x)
F_{\mu\nu}^{a}(x)\;, \quad 
\bar q(x_0)  =  \frac{1}{64\pi^2}\, \sum_{\vec x} 
\epsilon_{\mu\nu\rho\sigma}\, F_{\mu\nu}^a(x) F_{\rho\sigma}^a(x)\; ,
\eea
where the gluon field strength tensor
is the one in Eq.~(\ref{eq:Fmunu}) but with the trace removed. In particular
we focus on the expectation value
\be
C_{e} (x_0) = \frac{1}{L^3} \langle \bar e(x_0) \rangle\; ,
\ee
and on the correlators
\bea
C_{ee} (x_0,y_0) & = & \frac{1}{L^3} \langle \bar e(x_0)\, \bar e(y_0)  \rangle_c\; , \\
C_{qq} (x_0,y_0) & = & \frac{1}{L^3} \langle \bar q(x_0)\, \bar q(y_0)  \rangle\; . 
\eea
In an analysis of $200$ level-0 configurations each spaced by 8 MDUs, autocorrelations
of $\bar e (x_0)$ and $\bar q(x_0)$ are not detectable.

The two-level estimates of the same quantities have been carried out
by first averaging, for each of the $n_0$ configurations, the
densities over the $n_1$ level-1 background fields. This gives $n_0$   
measurements of the improved observables. The figure of merit is the variance of
this estimator. In the situation where autocorrelations among
the $n_0$ level-0 configurations can be neglected, the square root of the
variance divided by $\sqrt{n_0}$ gives the error of the measurement.
Since the cost of the simulation scales linearly in $n_1$, the variance
itself should decrease with $n_1$ to break even.

\begin{figure}
\begin{center}
\includegraphics[width=0.48\textwidth,clip]{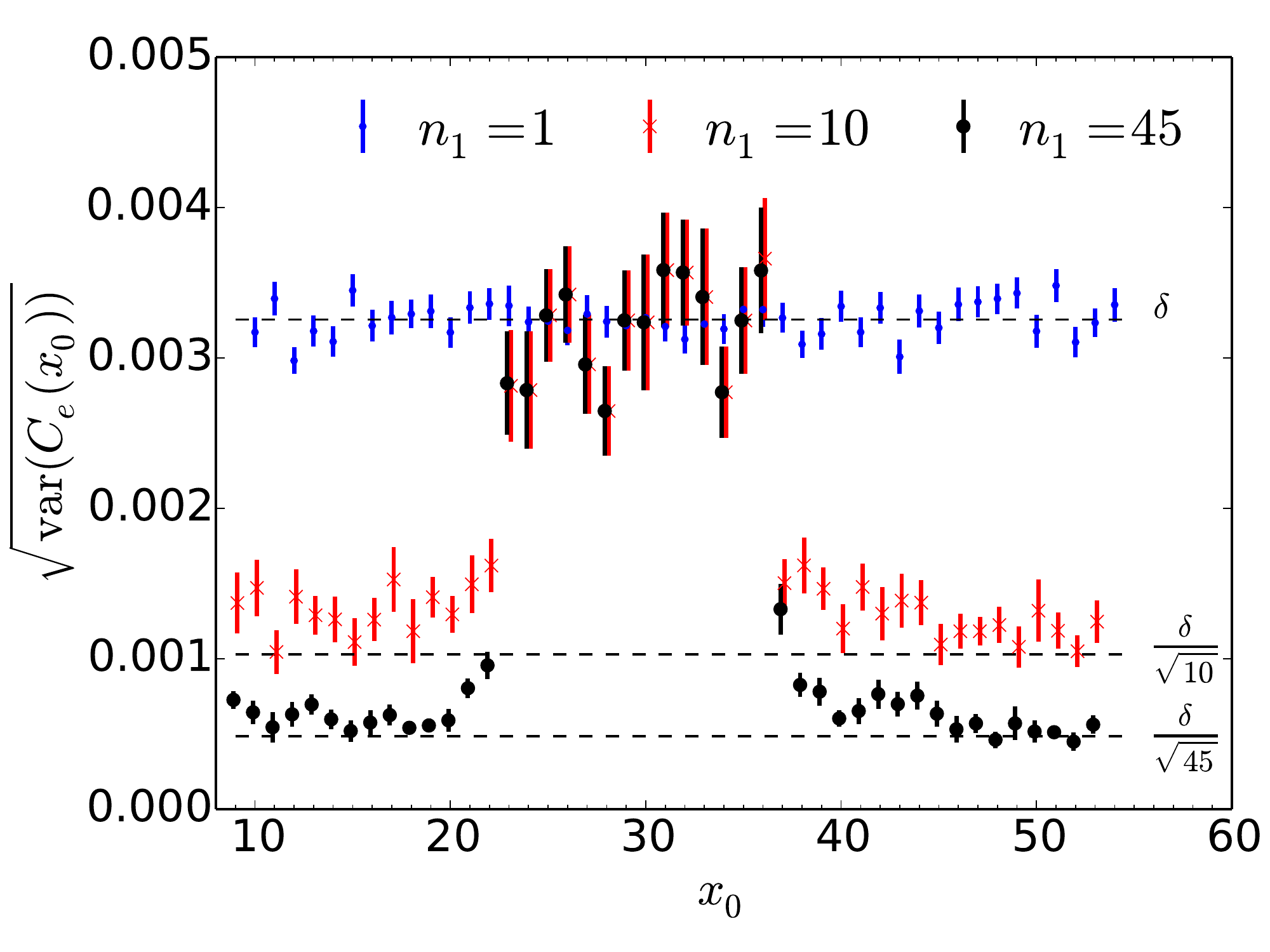}
\hspace{0.02\textwidth}
\includegraphics[width=0.48\textwidth,clip]{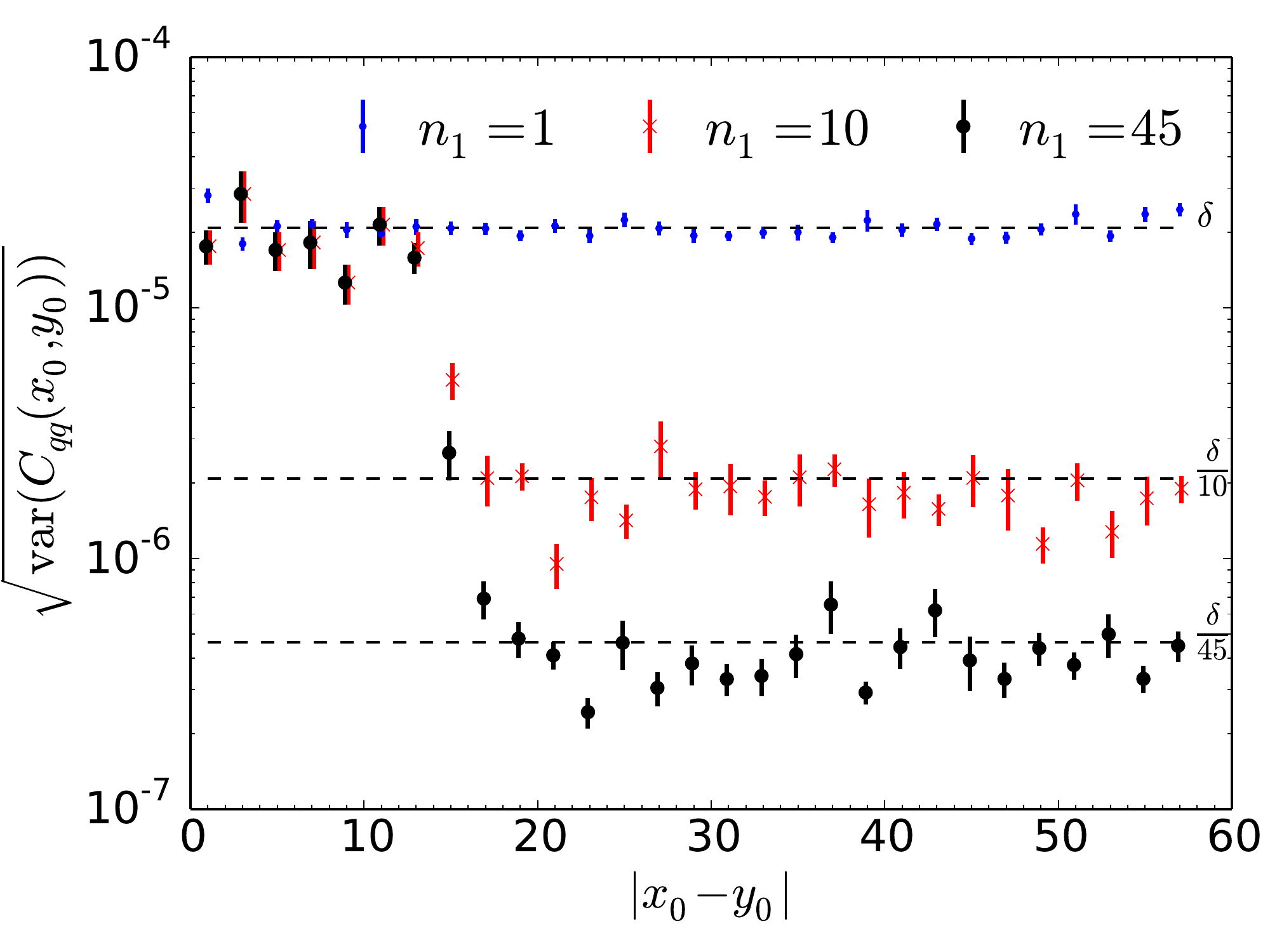}
\end{center}
\caption{\label{f:num}In the left panel, the square root of the variance of the energy density averaged 
over the time slice $x_0$ is shown. In the frozen central region this does not profit from
the level-1 updates, while in the active regions, it decreases with the square root of their inverse number.
The right plot demonstrates the effectiveness of the multilevel algorithm for the topological charge
density correlation function. The time slices $x_0$ and $y_0=30-x_0$ are chosen such that they are symmetric
with respect to the frozen region $\Lambda_1$. Once $|x_0-y_0|>12$ they enter the active regions where the
square root variance decreases with $1/n_1$. In both plots, the horizontal lines indicate the ideal scaling
behavior as expected from the variance measured at level 0.
}
\end{figure}

The square root of the variance of $C_{e} (x_0)$ as a function of $x_0$ is shown
in the left panel of Fig.~\ref{f:num}. In the central region, the links are
frozen during the level-1 updates. We therefore do expect the same 
variance as in the level-0 estimator, but the error of the variance is
larger due to the smaller value of $n_0$ in this case. Once we move into the
active regions $\Lambda_0$ and $\Lambda_2$, however, the variance of the 
estimator is clearly improved, in agreement with what is expected from 
ideal scaling, i.e.\ $\sqrt{\mathrm{var}(C_e)}\propto 1/\sqrt{n_1}$.

In the right panel the same analysis is shown for the two-point function
$C_{qq}$, and analogous results are obtained for $C_{ee}$.
Here the full benefit of the method can be realized, because an improved
estimator can be constructed by averaging for each of the $n_0$ fields the
densities in regions $\Lambda_0$ and $\Lambda_2$ 
independently before constructing the two-point function. As optimal
scaling in this case we expect a reduction of the square root of the variance, 
and therefore the error, with $1/n_1$. The numerical data are in agreement
with such a reduction once $x_0$ and $y_0$ are in two different active regions.

The picture emerging from this analysis is just in line with expectations. 
In the region where the links are
frozen during the level-1 updates no benefit from the multilevel is observed. As soon
as the densities are in the active regions $\Lambda_0$ and $\Lambda_2$, the square root of
the variances of the one- and two-point functions are reduced by $1/\sqrt{n_1}$ and
$1/n_1$ respectively. The two-level Monte Carlo works at full potentiality in these regions,
with a net gain in the computational cost of the two-point functions of $n_1$. This in turn
implies that links in the active regions $\Lambda_0$ and $\Lambda_2$ are regularly updated
during the level-1 MB-DD-HMC. In particular no freezing induced by multiboson fields is
observed.

\subsection{Disconnected pseudoscalar propagator\label{sec:etapN}}
Quark-line disconnected correlation functions serve as a second test of the method,
following our quenched results presented in \Ref{Ce:2016idq}.
We restrict ourselves to the correlation between two flavor-diagonal pseudoscalar densities,
\be
C_{P_d}(y_0,x_0) = \frac{1}{L^3}\, \langle\,
\sum_{{\vec x}} \tr\Big\{Q^{-1}(x,x)\Big\} \times \sum_{{\vec y}}
\tr\Big\{Q^{-1}(y,y)\Big\}\rangle
\label{e:disc}
\ee
which are decomposed as in Eq.~(6.1) of Ref.~\cite{Ce:2016idq}
\be\label{eq:splitsing}
C_{P_d}(y_0,x_0) = C^{(\text{f})}_{P_d}(y_0,x_0) + C^{(\text{r}_1)}_{P_d}(y_0,x_0) +
C^{(\text{r}_2)}_{P_d}(y_0,x_0)\; . 
\ee
In the first contribution, the two propagators in \Eq{e:disc} are replaced by approximate
propagators with Dirichlet boundary conditions imposed at
$x_0^{\rm cut}=30$, making this term amenable to multilevel integration. The other
two contributions are correction terms which make the equation exact, containing once or twice, respectively,
the difference between the full and approximate propagator. Note that in this case
there are several options for imposing Dirichlet boundary conditions. They may, for instance,
be imposed on the (opposite) respective ends of the frozen region $\Lambda_1$.

All the traces appearing on the rhs  of Eq.~(\ref{eq:splitsing})
are estimated stochastically by inverting the various Dirac operators on the
very same $n_\text{src}=500$ Gaussian random sources $\eta_i$, defined on the whole
space-time volume, and by contracting the solution with a time slice of $\eta_i$;
see Ref.~\cite{Ce:2016idq} for more details.\footnote{At variance with Ref.~\cite{Ce:2016idq},
here we did not use the hopping parameter variance reduction in the singlet evaluation.}
\begin{figure}[t!]
\centerline{\includegraphics[width=1.0\textwidth,clip]{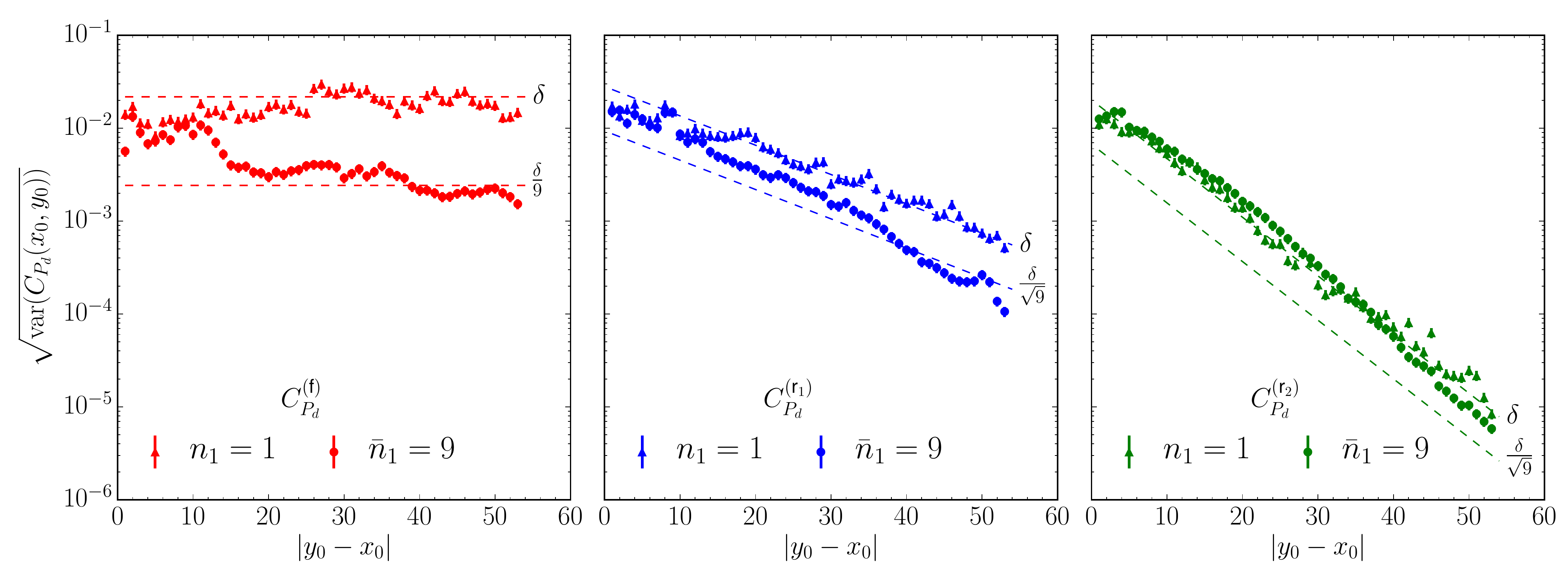}}
\caption{The square root of the variance of the three contributions
of the disconnected correlator of two flavor-diagonal pseudoscalar densities
[see Eq.~(\protect\ref{eq:splitsing})] are shown as a function of the time
separation $|y_0-x_0|$.
The dashed lines indicate the expected asymptotic form, which is constant
for the first contributions and falls exponentially $\exp\{-M_\pi |y_0-x_0|/2\}$ and
   $\exp\{-M_\pi |y_0-x_0|\}$ for the second and third contributions, respectively. In all
   panels, the lower line indicates the expected reduction from the level-1 updates.
\label{fig:singlet1}}
\end{figure}
A rough measure of the autocorrelation function of
$\sum_{{\vec x}} \tr\Big\{Q^{-1}(x,x)\Big\}$ from level-0 configurations shows that the
autocorrelation time is approximately $\tau_\mathrm{int}=10$ MDUs. The $32$
configurations can thus be safely considered as independent when used for level-0
measurements.

As for the gluonic two-point function above, the $C^{(\text{f})}_{P_d}(y_0,x_0)$ contribution,
for $x_0<24$ and $y_0>35$, is estimated by first averaging, for each of the
level-0 configurations, the two traces independently over the $n_1$
level-1 background fields. The product of the two means constitutes the improved estimator
which is then averaged over the $n_0$ measurements.
For the other two contributions such an improved estimator is not available. The correlation functions
$C^{(\text{r}_1)}_{P_d}$ and $C^{(\text{r}_2)}_{P_d}$ as a whole are first averaged over level 1, giving again 
$n_0$ measurements which are then processed in the usual manner.

While for the first contribution the scaling of the square root variance is expected to be proportional to $n_1^{-1}$,
for the second and the third it goes at most with $n_1^{-1/2}$. Since $n_1$ 
has to be counted in the number of independent measurements and
the level-1 updates are spaced by $4$~MDUs only,
in the plots we opt for normalizing the number of level-1 updates as
$\bar n_1 = n_1/(2\, \tau_\mathrm{int})=n_1/5$. This rescaling has no impact on the 
correctness of the procedure itself, but relates only to the level of improvement which one
can expect from a given number of level-1 updates.

The numerical results for the square root of the variance of 
$C^{(\text{f})}_{P_d}$, $C^{(\text{r}_1)}_{P_d}$, and $C^{(\text{r}_2)}_{P_d}$ are plotted in
Fig.~\ref{fig:singlet1} as a function of the time separation of the pseudoscalar
densities. In all cases  $x_0$ and $y_0$ belong to different domains, $y_0>x_0$,
and they are chosen to be as much as possible equidistant from $x_0^\text{cut}$.
These plots can be directly compared with those on the right column of Fig.~4 in
Ref.~\cite{Ce:2016idq}.

Our findings are very similar to the quenched
case~\cite{Ce:2016idq}, once one takes into account that in the present computation
$12$ slices are frozen instead of $1$. The square root of the variance of $C_{P_d}^{(\text{f})}$
(the left plot of Fig.~\ref{fig:singlet1}) is a flat function of $|y_0-x_0|$ with
sizable deviations near the boundaries of the domains. Up to the largest value that
we have, $\bar n_1=9$, the square root variance decreases approximately as
$\bar n_1^{-1}$ for large $|y_0-x_0|$; i.e.\ the
two-level Monte Carlo works as expected.
For $C_{P_d}^{(\text{r}_1)}$ and  $C_{P_d}^{(\text{r}_2)}$, center and right plots
in Fig.~\ref{fig:singlet1}, a strong dependence on $|y_0-x_0|$ is observed. They are compatible with an
exponential behavior of the form $\exp\{-M_\pi |y_0-x_0|/2\}$ and $\exp\{-M_\pi |y_0-x_0|\}$
respectively as suggested by theory.

The picture which emerges is analogous to the one in the quenched case~\cite{Ce:2016idq}. 
At large time distances, the statistical error of the standard estimate of the
disconnected pseudoscalar propagator is dominated by the one on $C_{P_d}^{(\text{f})}$. This is
the contribution for which the multilevel is efficient.
The second largest contribution is the statistical error on $C_{P_d}^{(\text{r}_1)}$ which, however,
is exponentially suppressed as $\exp\{-M_\pi |y_0-x_0|/2\}$. Therefore, once the two-level integration
is switched on and a large enough number of $n_1$ level-1 configurations is generated,
the signal-to-noise ratio increases exponentially in ${|y_0-x_0|}$
with respect to the standard Monte Carlo.

\section{Conclusions}
The gauge field dependence of the fermion determinant is factorizable by combining 
a domain decomposition with a multiboson representation of the (small) interaction
among gauge fields on distant blocks. The factorization does not require a particular
shape of the three domains, nor does each of them need to be connected.

The resulting action is local in the block scalar and gauge fields and can be simulated 
by variants of the standard hybrid Monte Carlo algorithm.  The measurements of 
local gluonic observables, such as the energy and the topological charge densities,
reveal a good efficiency of the algorithm in updating the gauge field. No particular
freezing of the links is observed.  The locality of the action can be beneficial for 
simulations using parallel computers due to reduced communication.

When combined with the recently proposed factorization of the fermion propagator~\cite{Ce:2016idq},
this setup naturally allows for multilevel Monte Carlo integration also in the presence of
fermions, opening new perspectives in lattice gauge theories. 
The numerical test of the disconnected correlator of two
flavor-diagonal pseudoscalar densities that we have reported indeed shows
that the signal-to-noise ratio increases exponentially with the time distance
of the sources when a two-level integration is at work.

Many interesting computations in lattice QCD and other theories
are expected to profit from these improvements, especially those which suffer from
signal-to-noise ratios which decrease exponentially with the time distance
of the sources. Prime examples here are disconnected correlators and/or
baryonic two- and three-point functions.

The  proposed method  relies on two key ingredients:
the locality of the Wilson Dirac operator and the (configuration by configuration)
exponential decrease of its inverse with the distance between the sink and the
source. The ideas and the
computational strategy presented here may, therefore, be applicable to 
very different theories with fermions if they enjoy these very basic properties. 

\section{Acknowledgments}
We thank M. L\"uscher for interesting discussions and comments to improve
the first version of this paper, in particular regarding the generality of the decomposition
proposed here and the simplification of Appendix \ref{a:invW}. Simulations have been
performed on the PC clusters
Galileo and Marconi
at CINECA (CINECA-INFN and CINECA-Bicocca agreements), PAX at DESY,
and Wilson at Milano-Bicocca. We thank these institutions for the computer
resources and the technical support. Numerical simulations have been carried
out with a modified version of the open-QCD code version
1.4~\cite{openQCD1.4}.

\appendix

\section{$O(a)$-improved Wilson-Dirac operator\label{app:Dw}}
The massive $O(a)$-improved Wilson-Dirac operator is defined
as~\cite{Sheikholeslami:1985ij,Luscher:1996sc,Luscher:2012av}
\be\label{eq:wdo}
D = D_\mathrm{w} + \delta D_\mathrm{v} + \delta D_\mathrm{b} + m_0\; ,
\ee
where $m_0$ is the bare quark mass, $D_\mathrm{w}$ is the massless Wilson-Dirac
operator
\begin{equation}
D_\mathrm{w}=
\frac{1}{2}\left\{ \gamma_\mu(\nabla^*_\mu+\nabla_\mu)- \nabla^*_\mu\nabla_\mu\right\}\; ,
\end{equation}
$\gamma_\mu$ are the Dirac matrices, and the summation over repeated indices is understood.
The covariant forward and backward derivatives 
$\nabla_\mu$ and $\nabla^*_\mu$ are defined to be
\be
\nabla_{\mu}\psi(x) =  U_\mu(x)\psi(x+\hat{\mu})-\psi(x)\; ,\quad
\nabla^*_\mu\psi(x) =  \psi(x)-  U_\mu^\dagger(x-\hat{\mu})
\psi(x-\hat{\mu})\; ,
\ee
where $U_\mu(x)$ are the link fields. The boundary correction terms are defined
to be 
\bea
\delta D_\mathrm{v}\psi(x) & = & c_{_{\rm SW}} \frac{i}{4} \sigma_{\mu\nu} \widehat F_{\mu\nu}(x)\psi(x)\; ,\\[0.25cm]
\delta D_\mathrm{b}\psi(x) & = &\{(c_{_{\rm F}} -1) \delta_{x_0,1} +
                                (c'_{_{\rm F}} -1) \delta_{x_0,T-1}\}\,\psi(x)\; , 
\eea
where with open boundary conditions in the time direction
$c'_{_{\rm F}}=c_{_{\rm F}}$, $\sigma_{\mu\nu}=\frac{i}{2}[\gamma_\mu,\gamma_\nu]$, and
the field strength of the gauge field is
\be\label{eq:Fmunu}
\widehat F_{\mu\nu}(x) = \frac{1}{8} \{Q_{\mu\nu}(x) -  Q_{\nu\mu}(x)\} 
\ee
with
\bea
Q_{\mu\nu}(x) & = & 
U_\mu(x)\, U_\nu(x+ \hat\mu)\, U^\dagger_\mu(x + \hat\nu)\,U^\dagger_\nu(x)\nonumber\\[0.125cm]
& + & U_\nu(x)\, U^\dagger_\mu(x-\hat\mu+\hat\nu)\, U^\dagger_\nu(x - \hat\mu)\,
U_\mu(x-\hat\mu)\\[0.125cm]
& + & U^\dagger_\mu(x-\hat\mu)\, U^\dagger_\nu(x-\hat\mu-\hat\nu)\, 
U_\mu(x-\hat\mu-\hat\nu)\,U_\nu(x-\hat\nu)\nonumber\\[0.125cm]
& + & 
U^\dagger_\nu(x-\hat\nu)\, U_\mu(x-\hat\nu)\, 
U_\nu(x + \hat\mu - \hat\nu)\, U^\dagger_\mu(x)\; . \nonumber
\eea
We are also interested in the operator $Q=\gamma_5 D$,
which is Hermitian since $D$ satisfies the $\gamma_5$-Hermiticity relation
$D = \gamma_5 D^\dagger \gamma_5$. 

\section{LU decomposition of a $2 \times 2$ block matrix\label{app:DD}}
A $2 \times 2$ block matrix can be decomposed as 
\be\label{eq:blkdec}
Q = 
\left(\begin{array}{c@{~~}c@{~~}}
Q_{\Gamma} & Q_{\partial\Gamma}\\[0.25cm]
Q_{\partial\Gamma^*} & Q_{\Gamma^*}
\end{array}   \right) =
\left(\begin{array}{c@{~~}c@{~~}}
I & \;\; Q_{\partial\Gamma}\, Q_{\Gamma^*}^{-1} \\[0.25cm]
0  & I
\end{array}   \right)
\left(\begin{array}{c@{~~}c@{~~}}
S_{\Gamma} & 0 \\[0.25cm]
Q_{\partial\Gamma^*}  & Q_{\Gamma^*} 
\end{array}   \right)\; ,
\ee
where the Schur complement is defined as
\be\label{eq:Schur}
S_{\Gamma} = Q_{\Gamma} - Q_{\partial\Gamma}\, Q_{\Gamma^*}^{-1}\, Q_{\partial\Gamma^*}\; . 
\ee
Its determinant can then be factorized 
\be\label{eq:detblock}
\det\, Q =\det \left(Q_{\Gamma} - Q_{\partial\Gamma}\, Q_{\Gamma^*}^{-1}\, Q_{\partial\Gamma^*}\right)\,
           \det Q_{\Gamma^*}\; , 
\ee
while the inverse is given by
\be\label{eq:Scmpt2}
Q^{-1} = 
\left(\begin{array}{c@{~~}c@{~~}}
S_{\Gamma}^{-1} &
-S^{-1}_{\Gamma} Q_{\partial\Gamma} Q_{\Gamma^*}^{-1}
\\[0.25cm]
- Q_{\Gamma^*}^{-1} Q_{\partial\Gamma^*} S^{-1}_{\Gamma} &
\;\; Q_{\Gamma^*}^{-1} + Q^{-1}_{\Gamma^*} Q_{\partial\Gamma^*} S^{-1}_{\Gamma} Q_{\partial\Gamma} Q^{-1}_{\Gamma^*}
\end{array}   \right)\; .
\ee
It is worth noting that $S^{-1}_{\Gamma}$ is the exact inverse of
$Q$ if the source and the sink positions are both in $\Gamma$.

\section{Even-odd block decomposition of the determinant\label{a:evenodd}}
The factorization described in Secs.~\ref{sec:BD} and \ref{sec:MBF}
can be generalized to the case of a lattice decomposed in many thick time slices.
The two-block partitioning in step 1 of Sec.~\ref{sec:BD} generalizes
to 
\be
\Gamma=\bigcup_\text{even $i$} \Lambda_i\; , \qquad \Gamma^*=\bigcup_\text{odd $i$} \Lambda_i\; ,
\ee
and the operator $Q_{\Omega^*_i}$ becomes the three-block matrix 
\begin{equation}
Q_{\Omega^*_i}=
\begin{pmatrix}
Q_{\Lambda_{i-1,i-1}} & Q_{\Lambda_{i-1,i}} &   0        \\ 
Q_{\Lambda_{i,i-1}}  & Q_{\Lambda_{i,i}}   & Q_{\Lambda_{i,i+1}}  \\
0                & Q_{\Lambda_{i+1,i}} & Q_{\Lambda_{i+1,i+1}} 
\end{pmatrix}\qquad i=2,4,\dots
\end{equation}
with the exception of the first ($i=0$) and the last (if the last block is even) matrices
which remain two-block operators.
Following the same four-step procedure of Sec.~\ref{sec:BD}, the determinant
of the Hermitian Dirac operator is factorized as 
\begin{equation}
\det Q =
\left( \prod_\text{odd $i$} \frac{1}{\det Q^{-1}_{\Lambda_{i,i}}} \right)
\left( \prod_\text{even $i$} \frac{1}{\det P_{\Lambda_{i}}\,  Q^{-1}_{\Omega^*_i}  \, P_{\Lambda_{i}}}\right) \det W_1\; ,
\end{equation}
where
\be
W_z\!\! =\!\! \begin{pmatrix}
    \ddots & \ddots \\
    \ddots & z P_{\partial\Lambda_{i-2}} & \!\!\!\!\! P_{\partial\Lambda_{i-2}} Q_{\Omega^*_{i-2}}^{-1}
    Q_{\Lambda_{i-1,i}} \\[0.25cm]
    & \!\!\!\!\!P_{\partial\Lambda_{i}} Q_{\Omega^*_i}^{-1} Q_{\Lambda_{i-1,i-2}} & z P_{\partial\Lambda_{i}}  &
      \!\!\!\!\! P_{\partial\Lambda_{i}} Q_{\Omega^*_i}^{-1} Q_{\Lambda_{i+1,i+2}} \\[0.25cm]
    & & \!\!\!\!\! P_{\partial\Lambda_{i+2}} Q_{\Omega^*_{i+2}}^{-1} Q_{\Lambda_{i+1,i}} & z P_{\partial\Lambda_{i+2}}  & \ddots \\[0.25cm]
    & & & \ddots & \ddots
  \end{pmatrix}\, .\!\!\!\!\!
\ee
The product $W^\dagger_z W_z$ has nonzero matrix elements on the diagonal and among
first and second nearest-neighbor even thick time slices. Each term in the corresponding
multiboson action, however, depends only on one three-block operator analogously to
Eq.~(\ref{eq:multiboson}); i.e.\ the dependence on the gauge field in the interior of the
even thick time slices is factorized. 

\section{Polynomial approximation of $1/z$ \label{app:b}}
The Chebyshev polynomials offer an (asymptotically) optimal
polynomial approximation of $1/z$ when $z$ is within an ellipse which
does not contain the origin, see Refs.~\cite{Manteuffel1977,Saad:2003}
and references therein.

When $z$ is contained in an ellipse
centered at a distance $d$ from the origin on the positive real axis,
with major and minor radii $a$ and $b$ respectively and with focus distance
$c=\sqrt{a^2-b^2}$, the polynomial approximation of $1/z$ of order $n$
is 
\begin{equation}
\label{eq:approx_poly}
P_N(z) \equiv \frac{1-R_{N+1}(z)}{z} = c_N \prod_{k=1}^N (z-z_k) \;,
\end{equation}
where
\begin{equation}
\label{eq:Rnp1}
R_{N+1}(z) \equiv \frac{T_{N+1}\left(\frac{d-z}{c}\right)}{T_{N+1}\left(\frac{d}{c}\right)} \;,
\end{equation}
with $T_{k}(z)$ being the Chebyshev polynomial of the first kind of degree $k$. The $N$ roots of
$P_N(z)$ are obtained by requiring that $R_{N+1}(0)=1$, and they are given by 
\begin{equation}\label{eq:zk}
z_k = d\left(1-\cos{\frac{2\pi k}{N+1}}\right) - i \sqrt{d^2-c^2} \sin{\frac{2\pi k}{N+1}}\; ,
\quad k=1,\dots,N  \;.
\end{equation}
They lie on the ellipse in the complex plane with center $d$, foci $d \pm c$, and which passes
through zero. By using the definition of the Chebyshev polynomials, a uniform error bound on
the approximation is given by
\be\label{eq:bound2}
|1- z P_N(z)|\!=\!|R_{N+1}(z)|\!\leq\! \left(\frac{a + \sqrt{a^2-c^2}}{d + \sqrt{d^2-c^2}}\right)^{\!\!N+1}
\!\!\!\!\left\{1 + \Big[\frac{a}{c} + \big (\frac{a^2}{c^2}-1\big)^{1/2}\Big]^{-2N-2}\right\}\; . 
\ee

\subsection{The circle}
In the limit $c\rightarrow 0$, when the ellipse becomes a circle centered
in $d$ with radius $a=b$, it holds
\be
|R_{N+1}(z)| \equiv \left|\frac{d-z}{d}\right|^{N+1}\; .
\ee
The bound becomes
\be
|1- z P_N(z)|=|R_{N+1}(z)|\leq \left(\frac{a}{d}\right)^{N+1}\; ,
\ee
which corresponds to the limit of Eq.~(\ref{eq:bound2}) when $c\rightarrow 0$.
The Zarantonello lemma guarantees that the polynomial is optimal
in this case. The roots of $P_N(z)$ are again given by Eq.~(\ref{eq:zk})
with $c=0$. They lie on a circle centered in $d$ of radius $d$.
If we choose $d+a=1$ and define the spectral gap as $d-a=\epsilon$ we get
\be
|R_{N+1}(z)|\leq \left(\frac{a}{d}\right)^{N+1} =
\left( \frac{1-\epsilon}{1+\epsilon} \right)^{N+1}\; .
\ee
In the limit in which $\epsilon\ll 1$
\be\label{eq:convC}
\left(\frac{1-\epsilon}{1+\epsilon} \right)^{N+1} \sim e^{-2 \epsilon (N+1)}\; , 
\ee
which shows that the polynomial approximation converges exponentially in $N$ with
a rate twice the (small) gap.

\section{Inverse of  $W_z$\label{a:invW}}
The operator $W_z$, defined in \Eq{eq:Wz}, acts on the union of the inner boundaries of regions
$\Lambda_0$ and $\Lambda_2$. For the heatbath of the bosonic fields, the solution to the
equation $W_z \chi = \eta$
needs to be found. This is not trivial, because the definition of $W_z$ itself contains
matrix inverses. Following the general idea of Eq.~(3.12) of \Ref{Luscher:2005rx}, we recast this 
problem into the solution of an extended system, from which a suitable projection gives the
desired result. In particular we define complex vectors on an extended lattice, the latter being
the ordinary lattice augmented by a copy of region $\Lambda_1$. On this space acts  the
matrix $\bar Q$, defined by
\begin{equation}
  \bar{Q} \equiv \begin{pmatrix} 
    z\, Q_{\Omega^*_0} & Q_{\Lambda_{1,2}}      \\[0.125cm]
    Q_{\Lambda_{1,0}} & z\, Q_{\Omega^*_1}
  \end{pmatrix} \; .
\end{equation}
This is an ordinary sparse matrix, which is amenable to the standard iterative algorithms for the solution
of linear systems. Since it is in general not well conditioned, it turns out to be profitable 
to use the deflation techniques introduced in \Ref{Luscher:2007se} to accelerate the computation.
By using the matrix $\bar Q$, it can be shown that the following identity holds
\begin{equation}
   W_z^{-1} 
   = \begin{pmatrix}  P_{\partial\Lambda_{0}} &0\\
     0&P_{\partial\Lambda_{2}} \end{pmatrix}\bar{Q}^{-1}
     \begin{pmatrix}
      Q_{\Omega^*_0}& 0  \\
      0 & Q_{\Omega^*_1} 
     \end{pmatrix}
     \begin{pmatrix}  P_{\partial\Lambda_{0}} &0\\
     0&P_{\partial\Lambda_{2}} \end{pmatrix}
 \,.
\end{equation}
The computation of the inverse of $W_z$ is thus reduced to the solution of a
sparse linear system.

\bibliographystyle{JHEP}
\bibliography{mb.bib}

\end{document}

%% file: macros.tex
\newcommand{\be}{\begin{equation}}
\newcommand{\ee}{\end{equation}}
\newcommand{\bea}{\begin{eqnarray}}
\newcommand{\eea}{\end{eqnarray}}
\newcommand{\bes}{\begin{eqnarray}}
\newcommand{\ees}{\end{eqnarray}}
\newcommand{\ba}{\begin{array}}
\newcommand{\ea}{\end{array}}

\newcommand{\Eq}[1]{Eq.~(\ref{#1})}

\newcommand{\Sect}[1]{Section~\ref{#1}}

\newcommand{\Ref}[1]{Ref.~\cite{#1}}
\newcommand{\Refs}[1]{Refs.~\cite{#1}}

\newcommand{\fm}{\mathrm{fm}}

%% file: paper.bbl
\providecommand{\href}[2]{#2}\begingroup\raggedright\begin{thebibliography}{10}

\bibitem{Duane:1987de}
S.~Duane, A.~D. Kennedy, B.~J. Pendleton, and D.~Roweth, {\it {Hybrid Monte
  Carlo}},  {\em Phys. Lett.} {\bf B195} (1987) 216--222.

\bibitem{Hasenbusch:1998yb}
M.~Hasenbusch, {\it {Speeding up finite step size updating of full QCD on the
  lattice}},  {\em Phys. Rev.} {\bf D59} (1999) 054505,
  [\href{http://arxiv.org/abs/hep-lat/9807031}{{\tt hep-lat/9807031}}].

\bibitem{Knechtli:2003yt}
{\bf Alpha} Collaboration, F.~Knechtli and U.~Wolff, {\it {Dynamical fermions
  as a global correction}},  {\em Nucl. Phys.} {\bf B663} (2003) 3--32,
  [\href{http://arxiv.org/abs/hep-lat/0303001}{{\tt hep-lat/0303001}}].

\bibitem{Hasenfratz:2005tt}
A.~Hasenfratz, P.~Hasenfratz, and F.~Niedermayer, {\it {Simulating full QCD
  with the fixed point action}},  {\em Phys. Rev.} {\bf D72} (2005) 114508,
  [\href{http://arxiv.org/abs/hep-lat/0506024}{{\tt hep-lat/0506024}}].

\bibitem{Gehrmann:1999wr}
B.~Gehrmann and U.~Wolff, {\it {Efficiencies and optimization of HMC algorithms
  in pure gauge theory}},  {\em Nucl. Phys. Proc. Suppl.} {\bf 83} (2000)
  801--803, [\href{http://arxiv.org/abs/hep-lat/9908003}{{\tt
  hep-lat/9908003}}].

\bibitem{Parisi:1983hm}
G.~Parisi, R.~Petronzio, and F.~Rapuano, {\it {A Measurement of the String
  Tension Near the Continuum Limit}},  {\em Phys. Lett.} {\bf B128} (1983)
  418--420.

\bibitem{Luscher:2001up}
M.~{L\"uscher} and P.~Weisz, {\it {Locality and exponential error reduction in
  numerical lattice gauge theory}},  {\em JHEP} {\bf 09} (2001) 010,
  [\href{http://arxiv.org/abs/hep-lat/0108014}{{\tt hep-lat/0108014}}].

\bibitem{Meyer:2002cd}
H.~B. Meyer, {\it {Locality and statistical error reduction on correlation
  functions}},  {\em JHEP} {\bf 01} (2003) 048,
  [\href{http://arxiv.org/abs/hep-lat/0209145}{{\tt hep-lat/0209145}}].

\bibitem{DellaMorte:2007zz}
M.~Della~Morte and L.~Giusti, {\it {Exploiting symmetries for exponential error
  reduction in path integral Monte Carlo}},  {\em Comput. Phys. Commun.} {\bf
  180} (2009) 813--818.

\bibitem{DellaMorte:2008jd}
M.~Della~Morte and L.~Giusti, {\it {Symmetries and exponential error reduction
  in Yang-Mills theories on the lattice}},  {\em Comput. Phys. Commun.} {\bf
  180} (2009) 819--826, [\href{http://arxiv.org/abs/0806.2601}{{\tt
  arXiv:0806.2601}}].

\bibitem{DellaMorte:2010yp}
M.~Della~Morte and L.~Giusti, {\it {A novel approach for computing glueball
  masses and matrix elements in Yang-Mills theories on the lattice}},  {\em
  JHEP} {\bf 05} (2011) 056, [\href{http://arxiv.org/abs/1012.2562}{{\tt
  arXiv:1012.2562}}].

\bibitem{Parisi:1983ae}
G.~Parisi, {\it {The Strategy for Computing the Hadronic Mass Spectrum}},  {\em
  Phys. Rept.} {\bf 103} (1984) 203--211.

\bibitem{Lepage:1989hd}
G.~P. Lepage, {\it {The Analysis of Algorithms for Lattice Field Theory}},  in
  {\em {Boulder ASI 1989:97-120}}, pp.~97--120, 1989.

\bibitem{Luscher:1993xx}
M.~{L\"uscher}, {\it {A New approach to the problem of dynamical quarks in
  numerical simulations of lattice QCD}},  {\em Nucl. Phys.} {\bf B418} (1994)
  637--648, [\href{http://arxiv.org/abs/hep-lat/9311007}{{\tt
  hep-lat/9311007}}].

\bibitem{Jegerlehner:1994cd}
B.~Jegerlehner, {\it {Study of a new simulation algorithm for dynamical quarks
  on the APE-100 parallel computer}},  {\em Nucl. Phys. Proc. Suppl.} {\bf 42}
  (1995) 879--881, [\href{http://arxiv.org/abs/hep-lat/9411065}{{\tt
  hep-lat/9411065}}].

\bibitem{Luscher:2005rx}
M.~{L\"uscher}, {\it {Schwarz-preconditioned HMC algorithm for two-flavour
  lattice QCD}},  {\em Comput. Phys. Commun.} {\bf 165} (2005) 199--220,
  [\href{http://arxiv.org/abs/hep-lat/0409106}{{\tt hep-lat/0409106}}].

\bibitem{Ce:2016idq}
M.~{C\`e}, L.~Giusti, and S.~Schaefer, {\it {Domain decomposition, multi-level
  integration and exponential noise reduction in lattice QCD}},  {\em Phys.
  Rev.} {\bf D93} (2016), no.~9 094507,
  [\href{http://arxiv.org/abs/1601.04587}{{\tt arXiv:1601.04587}}].

\bibitem{Radjavi:1969}
H.~Radjavi and J.~P. Williams, {\it Products of self-adjoint operators.},  {\em
  Michigan Math. J.} {\bf 16} (07, 1969) 177--185.

\bibitem{Borici:1995np}
A.~Borici and P.~de~Forcrand, {\it {Systematic errors of L\"uscher's fermion
  method and its extensions}},  {\em Nucl. Phys.} {\bf B454} (1995) 645--662,
  [\href{http://arxiv.org/abs/hep-lat/9505021}{{\tt hep-lat/9505021}}].

\bibitem{Borici:1995bk}
A.~Borici and P.~de~Forcrand, {\it {Variants of L\"uscher's fermion
  algorithm}},  {\em Nucl. Phys. Proc. Suppl.} {\bf 47} (1996) 800--803,
  [\href{http://arxiv.org/abs/hep-lat/9509080}{{\tt hep-lat/9509080}}].

\bibitem{Jegerlehner:1995wb}
B.~Jegerlehner, {\it {Improvements of {L\"uscher}'s local bosonic fermion
  algorithm}},  {\em Nucl. Phys.} {\bf B465} (1996) 487--506,
  [\href{http://arxiv.org/abs/hep-lat/9512001}{{\tt hep-lat/9512001}}].

\bibitem{Weingarten:1980hx}
D.~H. Weingarten and D.~N. Petcher, {\it {Monte Carlo Integration for Lattice
  Gauge Theories with Fermions}},  {\em Phys. Lett.} {\bf B99} (1981) 333--338.

\bibitem{Fritzsch:2012wq}
P.~Fritzsch, F.~Knechtli, B.~Leder, M.~Marinkovic, S.~Schaefer, R.~Sommer, and
  F.~Virotta, {\it {The strange quark mass and Lambda parameter of two flavor
  QCD}},  {\em Nucl. Phys.} {\bf B865} (2012) 397--429,
  [\href{http://arxiv.org/abs/1205.5380}{{\tt arXiv:1205.5380}}].

\bibitem{Hasenbusch:2001ne}
M.~Hasenbusch, {\it {Speeding up the hybrid Monte Carlo algorithm for dynamical
  fermions}},  {\em Phys.Lett.} {\bf B519} (2001) 177--182,
  [\href{http://arxiv.org/abs/hep-lat/0107019}{{\tt hep-lat/0107019}}].

\bibitem{Hasenbusch:2002ai}
M.~Hasenbusch and K.~Jansen, {\it {Speeding up lattice QCD simulations with
  clover improved Wilson fermions}},  {\em Nucl.Phys.} {\bf B659} (2003)
  299--320, [\href{http://arxiv.org/abs/hep-lat/0211042}{{\tt
  hep-lat/0211042}}].

\bibitem{Omelyan2003272}
I.~P. Omelyan, I.~M. Mryglod, and R.~Folk, {\it Symplectic analytically
  integrable decomposition algorithms: classification, derivation, and
  application to molecular dynamics, quantum and celestial mechanics
  simulations},  {\em Computer Physics Communications} {\bf 151} (2003), no.~3
  272 -- 314.

\bibitem{Luscher:2012av}
M.~L{\"u}scher and S.~Schaefer, {\it {Lattice QCD with open boundary conditions
  and twisted-mass reweighting}},  {\em Comput. Phys. Commun.} {\bf 184} (2013)
  519--528, [\href{http://arxiv.org/abs/1206.2809}{{\tt arXiv:1206.2809}}].

\bibitem{openQCD1.4}
 \href{http://arxiv.org/abs/{http://luscher.web.cern.ch/luscher/openQCD/}}{{\tt
  {http://luscher.web.cern.ch/luscher/openQCD/}}}.

\bibitem{Sheikholeslami:1985ij}
B.~Sheikholeslami and R.~Wohlert, {\it {Improved Continuum Limit Lattice Action
  for QCD with Wilson Fermions}},  {\em Nucl. Phys.} {\bf B259} (1985) 572.

\bibitem{Luscher:1996sc}
M.~{L\"uscher}, S.~Sint, R.~Sommer, and P.~Weisz, {\it {Chiral symmetry and
  O(a) improvement in lattice QCD}},  {\em Nucl. Phys.} {\bf B478} (1996)
  365--400, [\href{http://arxiv.org/abs/hep-lat/9605038}{{\tt
  hep-lat/9605038}}].

\bibitem{Manteuffel1977}
T.~A. Manteuffel, {\it {The Tchebychev iteration for nonsymmetric linear
  systems}},  {\em Numerische Mathematik} {\bf 28} (1977), no.~3 307--327.

\bibitem{Saad:2003}
Y.~Saad, {\em Iterative Methods for Sparse Linear Systems}.
\newblock Society for Industrial and Applied Mathematics, Philadelphia, PA,
  USA, 2nd~ed., 2003.

\bibitem{Luscher:2007se}
M.~{L\"uscher}, {\it {Local coherence and deflation of the low quark modes in
  lattice QCD}},  {\em JHEP} {\bf 07} (2007) 081,
  [\href{http://arxiv.org/abs/0706.2298}{{\tt arXiv:0706.2298}}].

\end{thebibliography}\endgroup
